\documentclass{emulateapj}

\shortauthors{HAYASHI \& CHIBA}
\shorttitle{Axisymmetric mass model for dSphs}

\begin{document}

\title{Probing non-spherical dark halos in the Galactic dwarf galaxies}

\author{Kohei~Hayashi\altaffilmark{1} and
	Masashi~Chiba\altaffilmark{1}}

\altaffiltext{1}{Astronomical Institute, Tohoku University,
Aoba-ku, Sendai 980-8578, Japan \\E-mail: {\it k.hayasi@astr.tohoku.ac.jp; chiba@astr.tohoku.ac.jp}}

\begin{abstract}
We construct axisymmetric mass models for dwarf spheroidal (dSph) galaxies in the Milky Way to obtain plausible limits on the non-spherical structure of their dark halos. This is motivated by the fact that the observed luminous parts of the dSphs are actually non-spherical and Cold Dark Matter (CDM) models predict non-spherical virialized dark halos. 
Our models consider velocity anisotropy of stars $\overline{v^2_R} / \overline{v^2_{\phi}}$, which can vary with the adopted cylindrical coordinates under the assumption $\overline{v^2_z}=\overline{v^2_R}$ for simplicity, and also include an inclination of the system as a fitting parameter to explain the observed line-of-sight velocity dispersion profile.
Applying these models to six of the bright dSphs in the Milky Way, we find that the best-fitting cases for most of the dSphs yield oblate and flattened dark halos, irrespective of assumed density profiles in their central parts. We also find that the total mass of the dSphs enclosed within a spheroid with major-axis length of 300~pc varies from $10^6M_{\odot}$ to $10^7M_{\odot}$, contrary to the conclusion from spherical models. This suggests the importance of considering shapes of dark halos in mass models of the dSphs. It is also found that dark halos of the Galactic dSphs may be more flattened than N-body predictions, thereby implying our yet incomplete understanding of baryonic and/or non-baryonic dark matter physics in dwarf galaxy scales.     

\end{abstract}


\section{INTRODUCTION}
Dwarf spheroidal (dSph) galaxies in the Milky Way are ideal sites for studying the basic properties of dark matter halos through their internal dynamics. This is because these satellites are sufficiently close that line-of-sight velocities for their resolved member stars can be measured by high-resolution spectroscopy (e.g., Kleyna et al. 2002; Walker et al. 2007; Mateo et al. 2008). Such spectroscopic observations have revealed that dSph galaxies have much larger velocity dispersions than expected from the stellar system alone, indicating that dSphs are largely dominated by dark matter, with mass-to-light ratios of 10 to 1000 (Mateo 1998; Gilmore et al. 2007). Moreover, these satellites have drawn special attention as building blocks of bright host galaxies within the framework of hierarchical structure formation theory, thereby providing fossil record in the evolution of the Galaxy and the Local Group (e.g., Tolstoy et al. 2009).

The $\Lambda$CDM models have played an indispensable role in describing such hierarchical formation  of galaxies in the $\Lambda$-dominated Universe, because the theory has well reproduced large-scale structure of galaxy distribution on spatial scales larger than $\sim 1$~Mpc. The process of structure formation is such that a number of small progenitors repeat merging and accretion, and aggregates into large objects in the growing process of self-gravitating structures. In recent years, advanced computational studies based on high-resolution N-body simulations have resolved important properties of dark matter halos at small spatial scales. First, a large number of dark matter substructures (subhalos) exist in a Milky Way-sized host halo (Moore et al. 1999; Klypin et al. 1999; Diemand et al. 2008; Springel et al. 2008). Second, all of these halos reveal their central densities being strongly cusped profiles (Navarro, Frenk \& White 1996, 1997; Fukushige \& Makino 1997; Moore et al. 1999; Diemand et al. 2008; Navarro et al. 2010) and their shapes being generally triaxial (Jing \& Suto 2000, 2002; Hayashi et al. 2007; Vera-Ciro et al. 2011; Allgood et al. 2006; Kuhlen et al. 2007; Schneider et al. 2011). 

However, these theoretical predictions are not in good agreement with observations, and it is yet a matter of ongoing debate.  For instance, the ``Cusp-Core'' problem is one of the open questions in $\Lambda$CDM theory: the central density profile of a dark halo is reported to be cored as suggested from observations of dSphs (Moore 1994; Burkert 1995; Gilmore et al. 2007) and Low Surface Brightness (LSB) galaxies (de Block et al. 2001; de Block \& Bosma 2002), whereas theoretically predicted central density is cuspy. Recent studies of this issue have claimed the possibility that the Galactic satellites can actually have a cusped density profile (Walker et al 2009c; Strigari et al. 2010; Battaglia et al. 2011). However, whether dSph galaxies are cusped or cored is yet unclear because of the presence of degeneracy in mass models.
Kinematic studies typically treat dSph galaxies as spherical symmetric systems with constant velocity anisotropy along the radii. However, in such models, there is a degeneracy between the velocity anisotropy of stars and various types of stellar and dark matter density profiles (Evans, An \& Walker 2009). Studies of dark matter in dSph galaxies have been hampered by this degeneracy. To overcome this ambiguity, at least in part, we need to consider more general models, where constancy of velocity anisotropy is relaxed. Another issue is that although the luminous parts of the dSphs are actually non-spherical with typical axial ratio of $0.6$ to $0.7$ (Irwin \& Hatzidimitriou 1995) and CDM models predict non-spherical virialized halos, most of mass models for dSphs have assumed spherical symmetry. Thus it is not able to derive and discuss shapes of dark halos in dSphs in comparison with theoretical predictions. 

 Motivated by the aforementioned problems, we construct non-spherical mass models for dSphs to obtain more realistic and important limits on density profiles and shapes of their dark halos. As a first step, we here work with axisymmetric mass models and related axisymmetric Jeans equations, where each of visible and halo density profiles has non-unity axial ratio. We also take into account a finite inclination angle of the system with respect to the line of sight. It is worth noting that such  axisymmetric models have often been used for bright elliptical galaxies (e.g., van der Marel et al. 1994; Magorrian \& Binney 1994; Cappellari 2008).
One of the benefits in our axisymmetric models is that not only the shapes of dark halos are derived from the fitting to observed velocity profiles, but also the change of velocity anisotropy with the spatial coordinates is fully taken into account.
  
 This paper is organized as follows. In \S 2, we describe our axisymmetric mass models and the method to solve axisymmetric Jeans equations to derive predicted line of sight velocity dispersions. In \S 3, we explain the kinematical and photometric data used for our work and the method of data analysis based on $\chi^2$ fitting. In \S 4, we present the results of $\chi^2$ fitting and compare with the results of spherical models.
Finally, in \S 5 we discuss halo mass estimated within inner 300 pc based on our model and implications for $\Lambda$CDM theory, and present our conclusions in \S 6.

\section{MODELS}
We construct axisymmetric mass models for dSphs, where each of luminous and dark matter density profiles has non-unity axial ratio, and solve axisymmetric Jeans equations to obtain projected velocity dispersion in the line of sight. We take into account a finite inclination angle of the system with respect to the line of sight.

\subsection{Stellar Density}
We assume that three-dimensional stellar densities of dSphs are modeled by those calculated from Plummer profiles (Plummer 1911), by which stellar surface densities of dSphs are commonly fit (e.g., Walker et al. 2009c). Here we generalize the corresponding three-dimensional stellar densities in the following axisymmetric form using cylindrical coordinates $(R,\phi,z)$:  
\begin{equation}
\nu(R,z) = \frac{3L}{4\pi b^3_{\ast}} \Bigl[1+\frac{m^2_{\ast}}{b^2_{\ast}}\Bigr]^{-5/2},
\label{eq:eq1}
\end{equation}
where $m^2_{\ast} = R^2 + z^2/q^2$, so $\nu$ is constant on ellipses with axial ratio $q$, and $L$ and $b_{\ast}$ are total luminosity and scale length, respectively. Surface densities of stars are now given as $I(x,y)=L(\pi b^2_{\ast})^{-1} (1+m^{\prime 2}_{\ast}/b^2_{\ast})^{-2}$ where $m^{\prime 2}_{\ast}=x^2+y^2/q^{\prime 2}$, $q^{\prime}$ is the projected axial ratio and $(x,y)$ are the coordinates aligned with the major and minor axes, respectively. Projected axial ratio $q^{\prime}$ is related to intrinsic ratio $q$ and inclination angle $i$ such as  $q^{\prime 2}=\cos^2i + q^2\sin^2i$, where $i=90^{\circ}$ when a galaxy is edge-on and $i=0^{\circ}$ for face-on. We estimate $b_{\ast}$ as the projected half light radius.

\begin{figure*}[t]
\begin{tabular}{ccc}
 \begin{minipage}{0.33\hsize}
  \begin{center}
   \includegraphics[width=65mm]{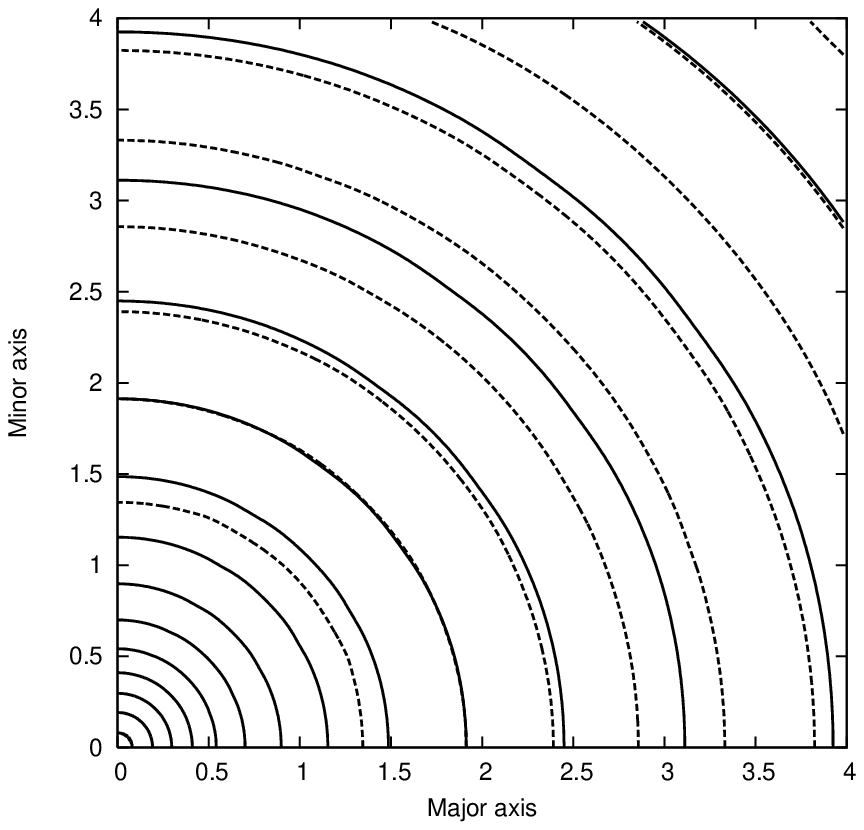}
  \end{center}
  \label{fig:one}
 \end{minipage}
  \begin{minipage}{0.33\hsize}
  \begin{center}
   \includegraphics[width=65mm]{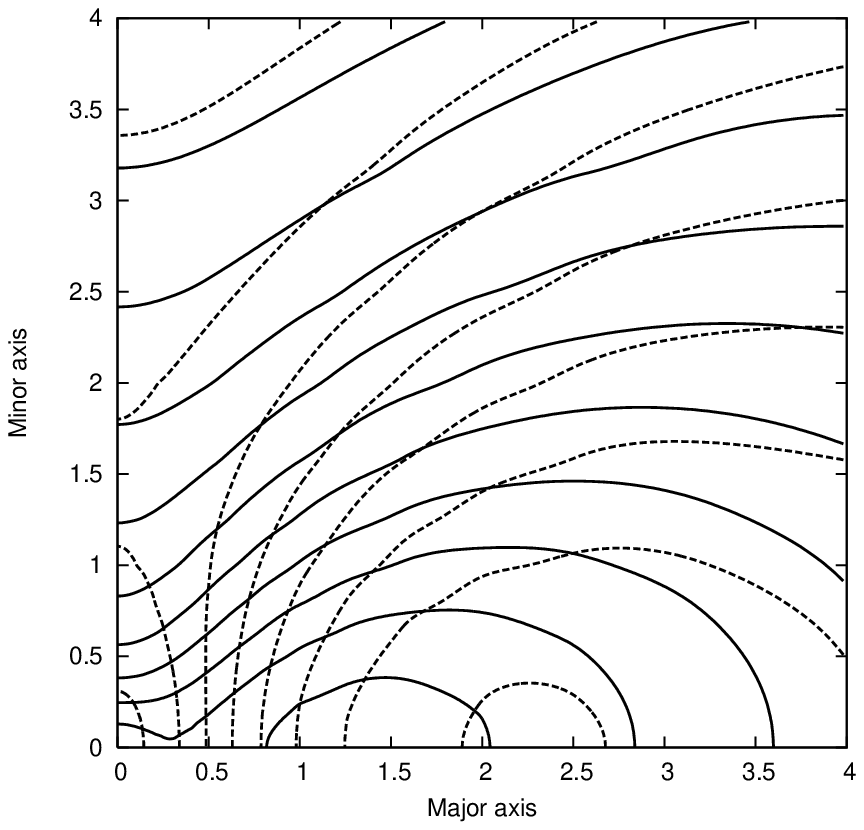}
  \end{center}
  \label{fig:two}
 \end{minipage}
  \begin{minipage}{0.33\hsize}
  \begin{center}
   \includegraphics[width=65mm]{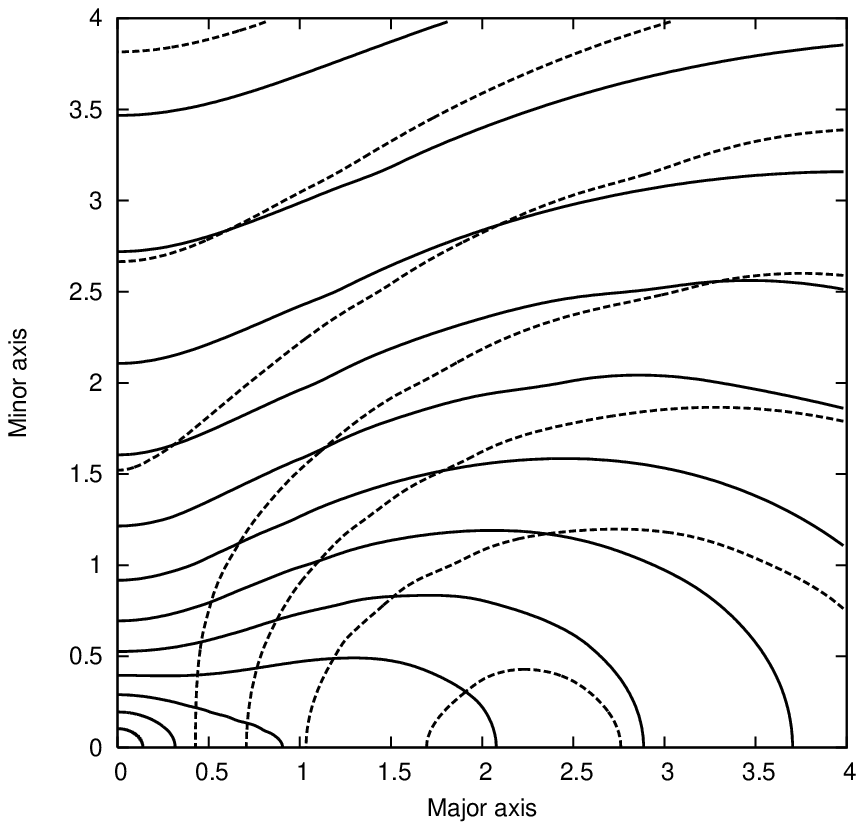}
  \end{center}
  \label{fig:three}
 \end{minipage}
 \end{tabular}
 \caption{Contours of line of sight velocity dispersion in the meridional plane derived from our axisymmetric models, where major and minor axis are normalized by a stellar scale length, i.e., $b_{*}$. We assume that the galactic inclination is edge-on (i.e., $i=90^{\circ}$) and the ratio of $b_{\rm halo}/b_{*}$ is unity. The solid line indicates NFW model ,while the dashed line is CORE model. Left panel shows when both of dark halo and luminous components are spherical $(Q=1, q=1)$. Middle panel shows the combination of a spherical dark halo $(Q=1)$ and non-spherical luminous part $(q=0.8)$, whereas right panel is for the case of $(Q=0.8, q=0.8)$.}
\label{fig:fig1}
\end{figure*}

\subsection{Halo Model}
 For the dark matter halo, we assume the following power-law form
 \begin{equation}
 \rho(R,z) = \rho_0 \Bigl(\frac{m}{b_{\rm halo}} \Bigr)^{\alpha}\Bigl[1+\Bigl(\frac{m}{b_{\rm halo}} \Bigr)^2 \Bigr]^{\delta},
 \label{eq:eq2}
 \end{equation}
\begin{equation}
m^2=R^2+z^2/Q^2,
\label{eq:eq3}
 \end{equation}
where $\rho_0$ is a scale density such that $\rho=2\rho_0$ at $m=b_{\rm halo}$ and $b_{\rm halo}$ is a scale length in the spatial distribution. In this work, we adopt four parameters ($Q, b_{\rm halo}, \rho_{0}$, $i$) for the halo model to be determined by fitting to the observed line of sight velocity dispersion. The model with $(\alpha,\delta)=(-1,-1)$ is well-known as the Navarro-Frenk-White profiles (hereafter ``NFW'': Navarro et al. 1997).  The NFW profiles have centrally cusped density and can reproduce cosmological N-body simulations well. On the other hand, those with $(\alpha,\delta)=(0,-1.5)$ have constant density cores and we call here as core profiles. These core profiles are suggested from  observations of dSph galaxies and LSB galaxies (e.g., Gilmore et al. 2007; de Block et al. 2001). In contrast to previous work, we set a new parameter $Q$, giving axial ratio of dark halos. For simplicity, we assume here that both stellar and halo's principal axes are aligned exactly. When viewed at inclination angle $i$, the projected isodensity contours are similar ellipses with axial ratio $Q^{\prime}$; $Q^{\prime 2}=\cos^2i +Q^2\sin^2i$.

The form of density profiles in equations (2) and (3) allows us to calculate the gravitational force in a simple manner (van der Marel et al. 1994; Binny \& Tremain 2008). Using variable constant, $\tau\equiv a^2_0e^2[\sinh ^2 u_m - (1/e-1)]$$(a_0 = const)$,  equation (\ref{eq:eq3}) is transformed to
\begin{equation}
\frac{m^2}{a^2_0} = \frac{R^2}{\tau + a^2_0} + \frac{z^2}{\tau + Q^2a^2_0}.
\label{eq:eq4}
\end{equation}
The gravitational force is thus given in the form of one dimensional integration:
\begin{equation}
{\bf g} = -\nabla\Phi = -\pi GQa_0\int^{\infty}_{0} d\tau\frac{\rho(m^2)\nabla m^2}{(\tau+a^2_0)\sqrt{\tau+Q^2a^2_0}},
\label{eq:eq5}
\end{equation} 
where
\begin{equation}
\nabla m^2 = 2a^2_0 \Bigl(\frac{R}{\tau + a^2_0}\hat{\bf e}_R + \frac{z}{\tau+Q^2a^2_0}\hat{\bf e}_z \Bigr),
\label{eq:eq6}
\end{equation}
and $(\hat{\bf e}_R , \hat{\bf e}_z)$ are unit vectors in the directions of $R$ and $z$, respectively. 

Mass interior to some distance $m^2=R^2+z^2/Q^2$ can be estimated in the following steps (See Binney \& Tremaine 2008). For the density in each shell being constant, mass of the shell between $m$ and $m + dm$ is given by 
\begin{equation}
\delta M = 4\pi\rho(m^2)\sqrt{1-e^2}m^2\delta m,
\label{eq:eq7}
\end{equation}
where $e$ is the eccentricity and $m$ is
\begin{equation}
m^2 = R^2 + \frac{z^2}{1-e^2}.
\label{eq:eq8}
\end{equation}
Hence $\sqrt{1-e^2}$ is equal to axial ratio $Q$ and equation (\ref{eq:eq7}) can be rewritten as 
\begin{equation}
\delta M = 4\pi\rho(m^2)Qm^2\delta m.
\label{eq:eq9}
\end{equation}
When we calculate this mass of spheroidal systems, we should integrate equation (\ref{eq:eq9}) from the mass center to arbitrary distance
\begin{equation}
M = \int^{m}_{0}4\pi\rho(m^2)Qm^2 dm.
\label{eq:eq10}
\end{equation}

\begin{figure*}[htbp]
 \begin{minipage}{0.5\hsize}
  \begin{center}
   \includegraphics[width=90mm]{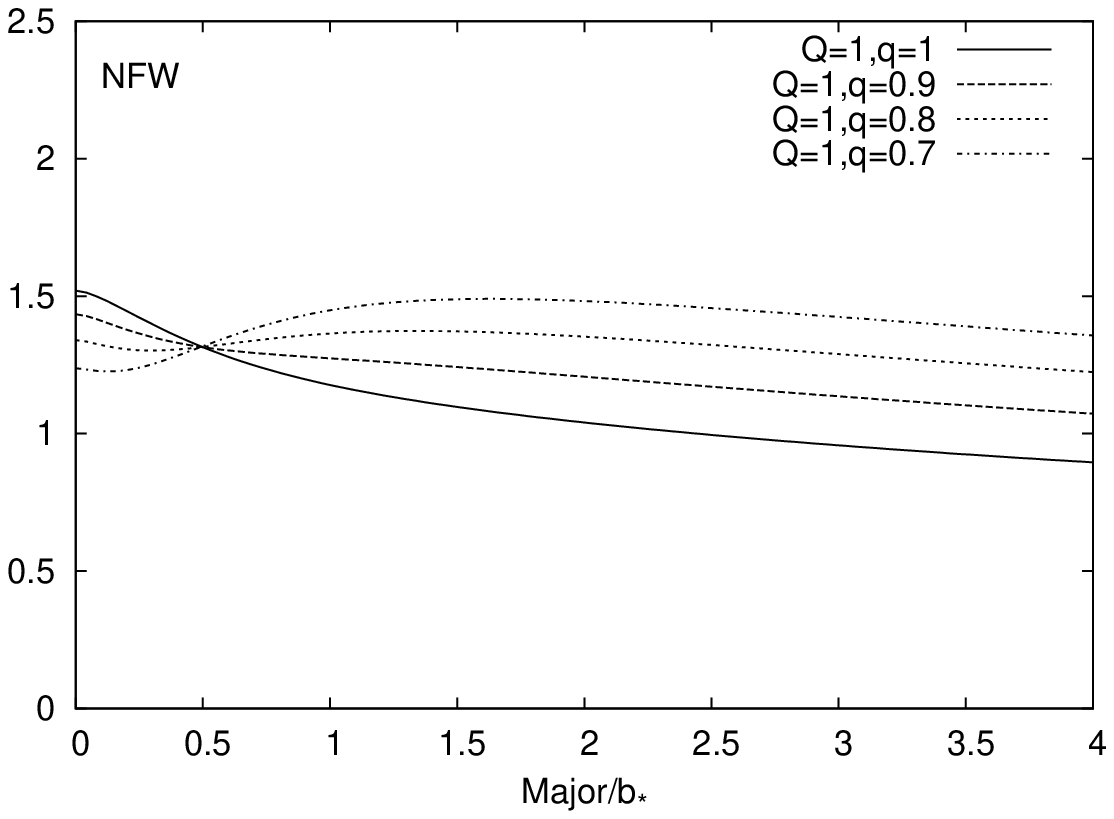}
  \end{center}
 \end{minipage}
 \begin{minipage}{0.5\hsize}
  \begin{center}
   \includegraphics[width=90mm]{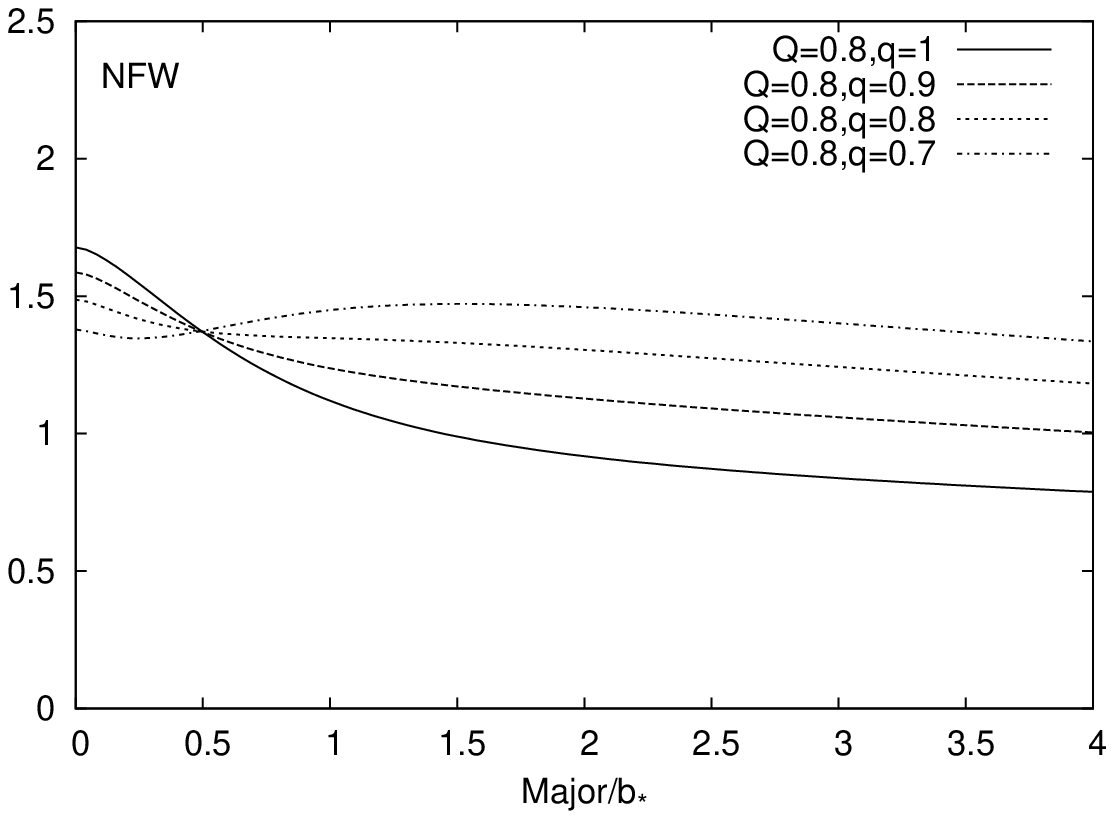}
  \end{center}
 \end{minipage}
  \begin{minipage}{0.5\hsize}
  \begin{center}
   \includegraphics[width=90mm]{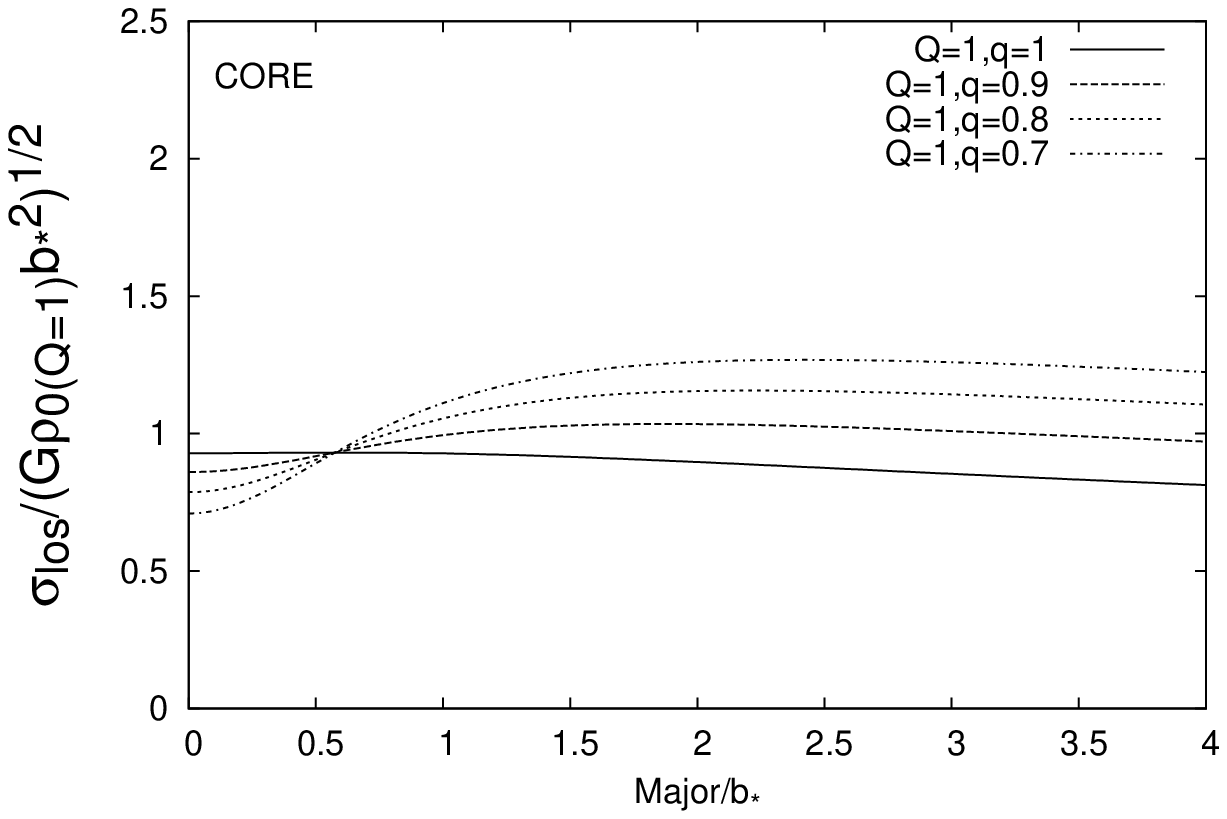}
  \end{center}
 \end{minipage}
 \begin{minipage}{0.5\hsize}
  \begin{center}
   \includegraphics[width=90mm]{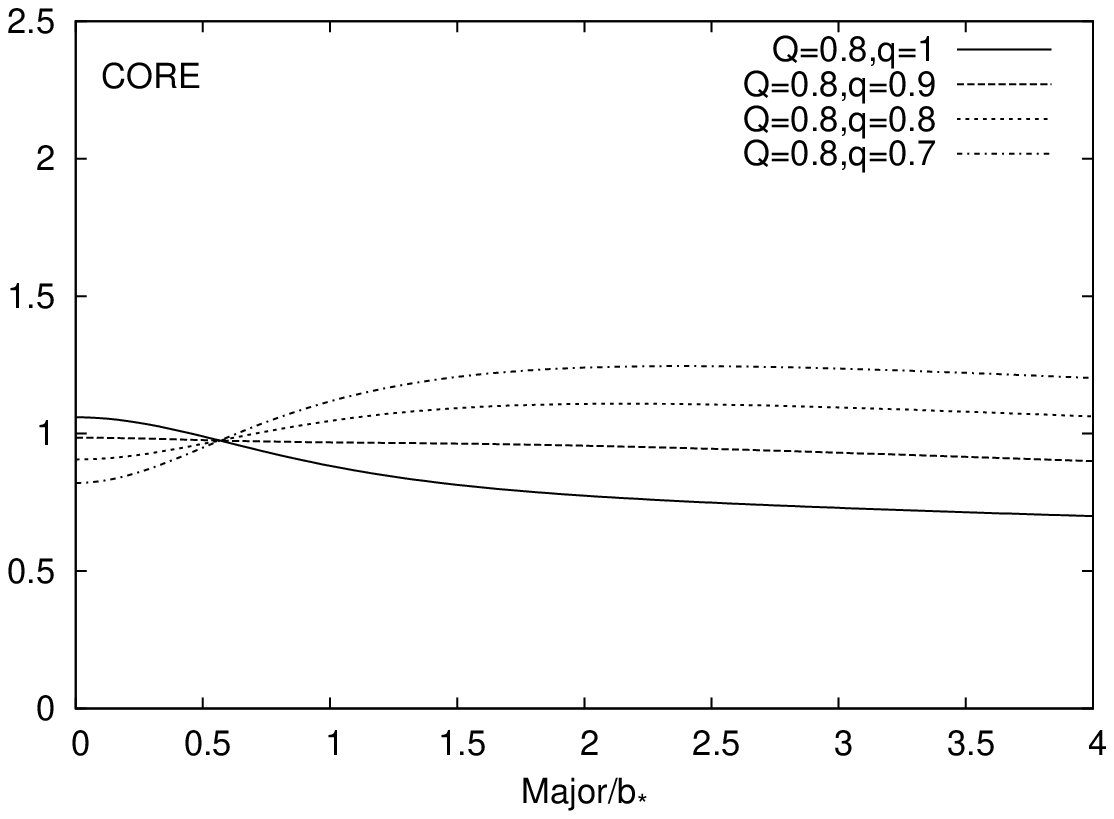}
  \end{center}
 \end{minipage}
  \begin{minipage}{0.5\hsize}
  \begin{center}
   \includegraphics[width=90mm]{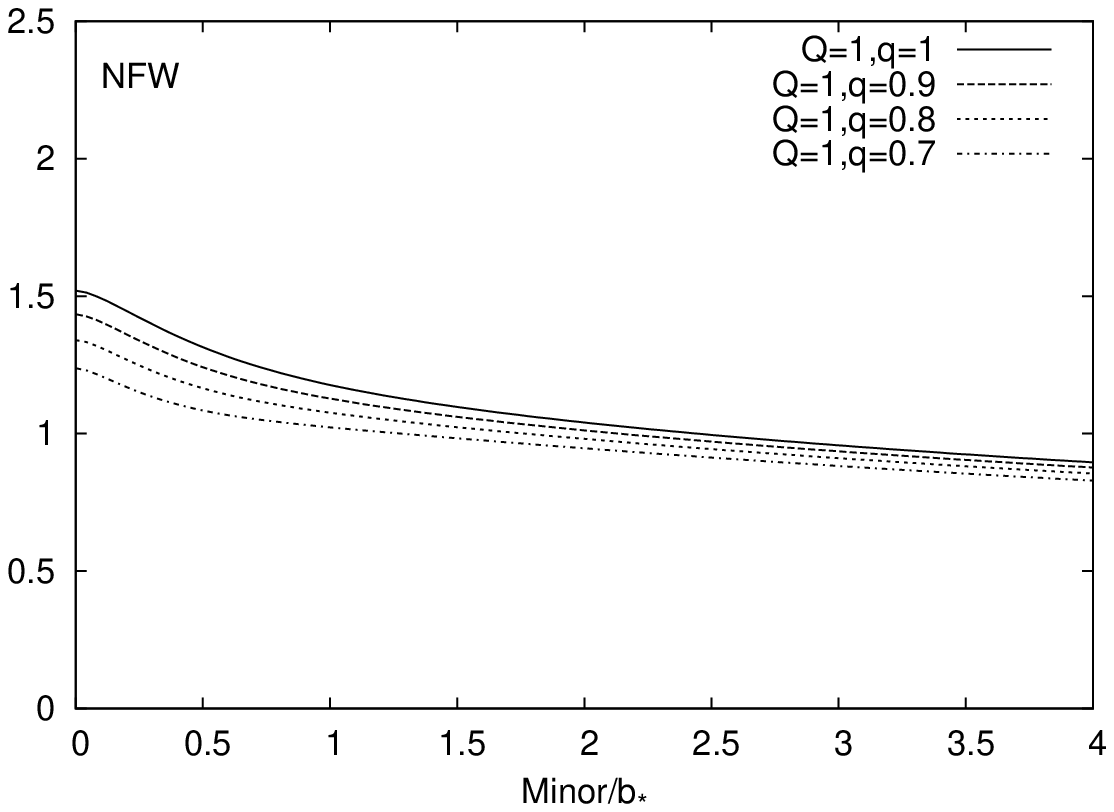}
  \end{center}
 \end{minipage}
 \begin{minipage}{0.5\hsize}
  \begin{center}
   \includegraphics[width=90mm]{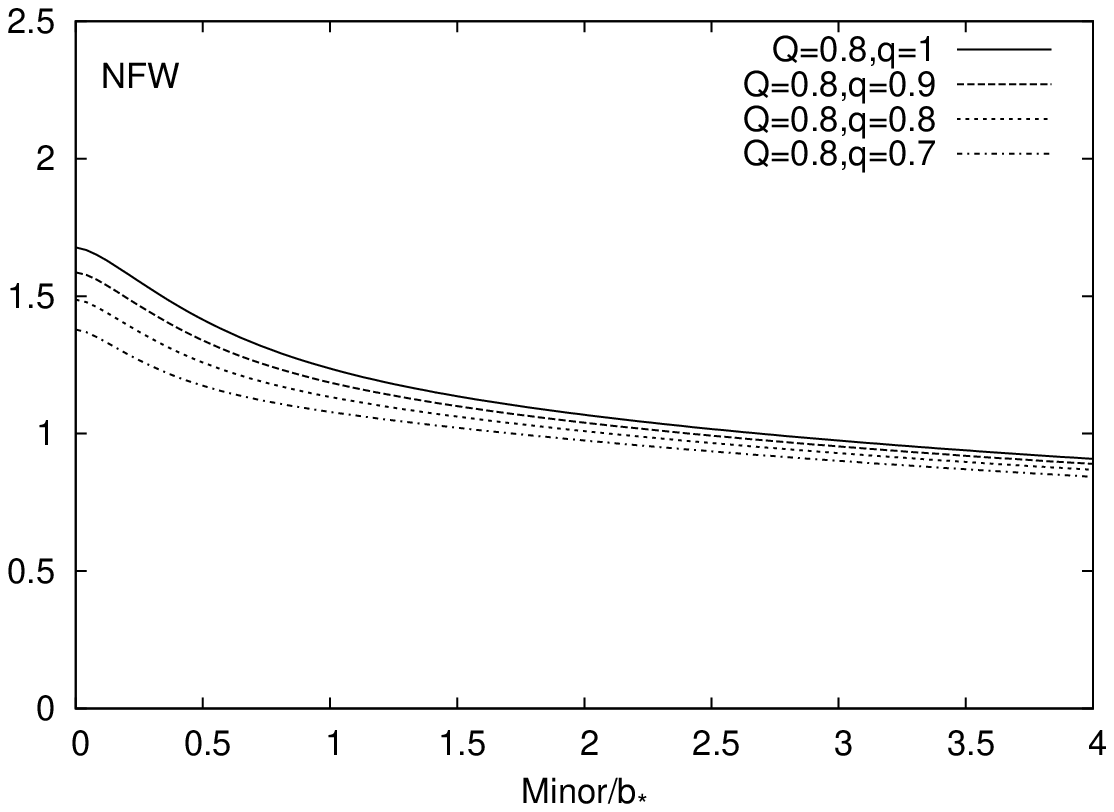}
  \end{center}
 \end{minipage}
 \caption{The upper panels show normalized line of sight velocity dispersions, $\sigma_{\rm los}/(G\rho_{0(Q=1)}b^2_{\ast})^{1/2}$, along the major axis for NFW, whereas the middle panels are for core density profiles. The bottom panels show these velocity dispersions along the minor axis for NFW. For all of these cases we assume that the inclination of a galaxy is edge-on ($i=90^{\circ}$)and the ratio of $b_{\rm halo}/b_{\ast}$ is unity for the sake of demonstration.}
  \label{fig:fig2}
 \end{figure*}

\subsection{Axisymmetric Jeans equations}

We assume that the stellar component of dSphs is in dynamical equilibrium with a gravitational potential dominated by dark matter. For the velocity distribution of stars, we assume that radial velocity dispersion, $\overline{v^2_{R}}$, is equal to that in $z$ direction, $\overline{v^2_z}$, which is equivalent to the assumption that a stellar distribution function is given as $f(E,L_z)$, i.e., depending on energy $E=\frac{1}{2}v^2+\Phi$ and angular momentum component $L_z=Rv_{\phi}$. $\Phi$ is a gravitational potential and $v_{\phi}$ is an azimuthal velocity component. Thus axisymmetric Jeans equations are written as
\begin{eqnarray}
\frac{\partial \nu \overline{v^2_z}}{\partial z}  &+& \nu\frac{\partial \Phi}{\partial z}= 0, 
\label{eq:eq11}\\
\frac{\partial \nu \overline{v^2_R}}{\partial R} &+& \nu\frac{\partial \Phi}{\partial R}
+ \frac{\nu(\overline{v^2_R} - \overline{v^2_{\phi}})}{R} = 0,
\label{eq:eq12}
\end{eqnarray}
where $\nu$ is the stellar density given in equation (\ref{eq:eq1}). Provided that $\Phi$ is dominated by a background dark halo, equation (\ref{eq:eq11}) can be integrated to yield 
\begin{equation}
\overline{v^2_z} = \overline{v^2_R} = \sigma^2 = \frac{1}{\nu(R,z)}\int^{\infty}_z \nu\frac{\partial \Phi}{\partial z}dz.
\label{eq:eq13}
\end{equation}
Now that $\overline{v^2_R}$ is known, we can obtain $\overline{v^2_{\phi}}$ from equation $(\ref{eq:eq12})$ :
\begin{equation}
\overline{v^2_{\phi}} = \sigma^2 + R \frac{\partial \Phi}{\partial R} + \frac{R}{\nu}\frac{\partial(\nu\sigma^2)}{\partial R}.
\label{eq:eq14}
\end{equation}

In order to compare these solutions with observed stellar kinematics in dSphs, we derive the line of sight velocity dispersion from $\sigma^2$ and $\overline{v^2_{\phi}}$, taking into account inclination between the line of sight and the galactic plane. We adopt the following steps for this calculation, following the method given in Tempel \& Tenjis (1990). Firstly, we project $\overline {v^2_R}$ ($=\sigma^2$)  and $\overline{v^2_{\phi}}$ to the plane parallel to the galactic plane. Projected dispersions are given as 
\begin{equation}
\sigma^2_{\ast} = \overline{v^2_{\phi}}\frac{x^2}{R^2} + \sigma^2\Bigl(1-\frac{x^2}{R^2} \Bigr),
\label{eq:eq15}
\end{equation}
where $x$ is the projected coordinate. Secondly, we project $\overline{v^2_{z}}$ ($=\sigma^2$) and $\sigma^2_{\ast}$ to the line of sight. Using $\Theta$ as the angle between the galactic plane and the line of sight, which is expressed by $\Theta = i - 90^{\circ}$, the line-of-sight velocity dispersion is 
\begin{equation}
\sigma^2_{\ell} = \sigma^2_{\ast}\cos^2\Theta + \sigma^2\sin^2\Theta.
 \label{eq:eq16}
\end{equation}
Finally, we average $\sigma^2_{\ell}$ along the line of sight by means of weighted integration with stellar density, thus 
\begin{equation} 
\sigma^2_{\rm los}(x,y) = \frac{1}{I(x,y)}\int^{\infty}_{-\infty} \nu(R,z)\sigma^2_{\ell}(R,z)d\ell,
\label{eq:eq17}
\end{equation}
where $I(x,y)$ is the surface density as determined from $\nu(R,z)$, and $\ell$ is defined along the line of sight.

\subsection{Model properties}

In order to demonstrate the impact of the non-spherical shape of stellar and dark matter components on line-of-sight velocity dispersion profiles, Figure \ref{fig:fig1} shows the predicted two dimensional distribution of $\sigma_{\rm los}$ calculated from our axisymmetric models for NFW (solid line) and cored (dashed line) dark matter halos, where we set $b_{\rm halo}/b_{\ast} = 1$ and $\rho_0=1$. Left panel shows spherical symmetry models, namely both of axial ratios of dark halos, $Q$, and luminous parts, $q$, are unity where we assume isotropic velocity dispersions for the sake of demonstration. Obviously, the model reproduces the result of the spherically symmetric mass model. On the other hand, middle and right panels show the predictions of non-spherical models. While middle panel shows the combination of a spherical dark halo ($Q=1$) and a non-spherical luminous part ($q=0.8$), in right panel both components are non-spherical ($Q=0.8, q=0.8$).  It is clear that these axisymmetric models show very different velocity dispersion profiles from the case of spherically symmetric models showing spherically symmetric kinematics of stars. Thus, these examples suggest that non-spherical matter distribution can be deduced from such characteristic distributions of stellar kinematics.

We here explain the effect of changing axial ratios of a stellar system, $q$, and dark halo, $Q$, on velocity dispersion profiles, $\sigma_{\rm los}$ (See Figure \ref{fig:fig2}). First, the effect of decreasing $q$ from unity while $Q$ is fixed, i.e., a more flattened stellar system, yields trough and crest-like features in central and outer parts of the $\sigma_{\rm los}$ profile, respectively, along the major axis, thereby showing the wavy $\sigma_{\rm los}$ profile (as seen in upper-left and middle-left panels of Figure 2). This is due to the decrease of $\sigma^2 (=\overline{v^2_{R}}=\overline{v^2_z})$ in central parts at $z=0$ as deduced from equation (\ref{eq:eq13}), so $\sigma_{\rm los}$ is reduced in such regions, while $\overline{v^2_{\phi}}$ is more dominant and thus $\sigma_{\rm los}$ is larger in outer parts, following equation (\ref{eq:eq14}). In much outer parts, $\sigma_{\rm los}$ is decreasing with radius as gravitational force becomes weaker at such large distances. 
Second, the effect of decreasing $Q$ from unity while $q$ is fixed, i.e., a more flattened dark halo, weakens the above wavy feature of the $\sigma_{\rm los}$ profile caused by non-unity $q$ along the major axis (as seen in upper-right and middle-right panels of Figure 2). This is because non-unity $Q$ yields larger gravitational force in $z$-direction, thereby increasing $\sigma^2$ in inner parts. Thus, the effect of decreasing $Q$ is opposed to that of decreasing $q$, thereby making the $\sigma_{\rm los}$ profile being a rather flat feature along the major axis.
On the other hand, along the minor axis, the effect of decreasing $q$ and $Q$ on $\sigma_{\rm los}$  is monotonous (as seen in bottom panels of Figure 2), which can be straightforwardly understood. From equation (\ref{eq:eq14}), $\overline{v^2_{\phi}}$ is equal to $\sigma^2$ along the minor axis of $R=0$, thus we can consider only the effect on $\sigma^2$, which is already described above: decreasing $q$ $(Q)$ reduces (increases) $\sigma^2$ and thus flattens (steepens) the $\sigma_{\rm los}$ profile.
Other halo parameters, i.e., a halo scale length, $b_{\rm halo}$, and scale density, $\rho_0$, mainly affect the amplitude of the $\sigma_{\rm los}$ profile and only weakly change its overall shape. Thus, $Q$ is insensitive to these parameters.
Besides, assumed inner slopes of dark halo densities also affect the central parts of the $\sigma_{\rm los}$ profile. In comparison with core profiles, NFW profiles have steeper inner slopes, thus the innermost part of the $\sigma_{\rm los}$ profile increases for the case of NFW ones. Therefore we are able to understand the difference between top and middle panels of Figure \ref{fig:fig2}. 

To compare these model results with observations as detailed below, we confine ourselves to $\sigma_{\rm los}$ profiles along the major, minor and intermediate axis, where the latter axis is defined at position angle of $45^{\circ}$ from the major axis. The choice of these axes is because predicted $\sigma_{\rm los}$ profiles along these reveal characteristic features, which are sufficiently different to discriminate different mass models. 

\begin{deluxetable*}{cccccccc}[t!!]
\label{tb:tab1}
\tablecolumns{8}
\tablewidth{6.0in}
\tablecaption{The observational dataset for six dSph satellites}
\tablehead{
Object & Number of stars & $L_V$($\times10^6L_{\odot}$)\tablenotemark{a} & $M_V$\tablenotemark{a}  &P.A.(deg)\tablenotemark{c} & distance (kpc)\tablenotemark{c} & $r_{\rm half}$(pc)\tablenotemark{b} &  $q^{\prime}$(axial ratio)\tablenotemark{c}}
\startdata
Carina & 776 & $0.24 \pm 0.1$ & $-8.6 \pm 0.5$ & $65\pm5$ &$85\pm5$& $241\pm 23$  & $0.67\pm0.05$\\
Fornax & 2523 & $14.0 \pm 4.0$ &  $-13.0 \pm 0.3$ & $41\pm1$& $120\pm8$ & $668\pm 34$ & $0.70\pm0.01$\\
Sculptor & 1360 & $1.4 \pm 0.6$ & $-10.7 \pm 0.5$& $99\pm1$ & $72\pm5$ & $260\pm 39$ & $0.68\pm0.03$\\
Sextans & 445 & $0.4 \pm 0.2$ & $-9.2 \pm 0.5$ & $56\pm5$ & $83\pm9$ & $682\pm 117$ & $0.65\pm0.05$\\
Draco & 185 & $0.18\pm 0.08$ & $-8.3 \pm 0.5$ & $89\pm2$& $76\pm5$ & $196\pm 12$ & $0.69\pm0.02$\\
LeoI & 328 & $3.4 \pm 1.1$ & $-11.5 \pm 0.3$ & $79\pm3$& $198\pm30$ & $246\pm 19$ & $0.79\pm0.03$
\enddata
\tablenotetext{a}{Taken from Irwin \& Hatzidimitriou (1995).}
\tablenotetext{b}{These values have been derived by Walker et al. (2009c) using Irwin \& Hatzidimitriou (1995) for Carina, Fornax, Sculptor, Sextans and Leo~I and by Martin et al. (2008) for Draco.}
\tablenotetext{c}{Taken from Irwin \& Hatzidimitriou (1995) except for Draco (Martin et al. 2008).}
\end{deluxetable*}

\section{DATA ANALYSIS}
In this section, we briefly describe photometric and kinematic data of member stars in six of the bright Galactic dSphs (Carina, Fornax, Sculptor, Sextans, Draco and Leo~I) for the application of our mass models. The method in fitting model predictions to the observed velocity data is also presented.

\subsection{Fundamental data}
We apply these axisymmetric models to the above-mentioned six dSphs to obtain their halo parameters ($Q, b_{\rm halo}, \rho_0$, $i$). For the kinematic data of their member stars, we adopt the following literatures. For Carina, Fornax, Sculptor and Sextans dSphs, we use published data in Walker et al. (2009a; 2009b), and for Draco and Leo~I, we adopt Kleyna et al. (2002) and Mateo et al. (2008), respectively.
Table 1 shows the observed properties of six dSph satellites: number of member stars, total V-band luminosity, V-band absolute magnitude, position angle, distance from the Galactic center, projected half light radius and axial ratio of the stellar system. Photometric data listed in column 3, 4 and 8 are adopted from Irwin \& Hatzidimitriou (1995). As previously mentioned, we adopt projected half light radius (column 7 in Table 1) as stellar scale length, $b_{\ast}$, in our models, for which we use the estimate in Walker et al. (2009) based on the Irwin \& Hatzidimitriou (1995) data. Also we use the observed axial ratio (column 8 in Table 1) as projected axial ratio $q^{\prime}$. 

\subsection{Velocity dispersion profiles}
The kinematic data sets that we use here are line-of-sight stellar velocities taken from the above cited papers. The method for evaluating membership and removing contaminations differs in each paper. First, for Carina, Fornax, Sculptor and Sextans dSphs, we use the result of an 'expectation-maximization' method from Walker et al. (2009b). For Draco, the separation of the member stars from the Galactic contaminant stars is clearly made so there is little likelihood of non-Draco stars being included in the samples. For Leo~I member stars, we use Mateo et al. (2008), which consider those stars that have velocities in the range from 240 to 320 ${\rm km}$ ${\rm s^{-1}}$. This range of velocities is well separated from the Galactic foreground stars, because it is unlikely that these contaminations present in this velocity range. The column 2 in Table 1 shows the number of member stars from which each method distinguishes contamination. These resolved stars are red giant branch stars that are straightforward to identify. 

In order to estimate the line-of-sight velocity dispersion profiles for each satellite, we adopt here the standard approach of using binned profiles. In particular, we focus on velocity dispersion profiles along three angles of axis: major axis, minor axis and intermediate axis which is defined at $45^{\circ}$ from the  major axis. 
Since we assume axisymmetry in this work, we analyze the velocity data along each axis by folding up the stellar distribution into the first quadrant for each of six dSphs. We then set boxes with 100~pc~$\times$~length $L$, where the former side is perpendicular to each axis in concern and the latter side (with length $L$) defined along the axis is set so that the nearly equal number of stars is contained in each box: $\sim 100$ stars/box for Carina, Fornax and Sculptor, $\sim 50$ stars/box for Sextans, and $\sim 25$ stars/box for Draco and Leo~I. We thus derive the velocity dispersion profiles by using the velocity data of stars contained in each box as defined above. This method is in contrast to previous works, where binned circular annuli are used based on the assumption of spherical symmetry.
Figures \ref{fig:fig3} and \ref{fig:fig4} display the analysis results of velocity dispersion profiles for six dSph satellites. It is found that velocity dispersion profiles along each axis show some systematic change from the galactic center. Therefore, these profiles allow us to derive the properties of non-spherical dark matter halos. To obtain four halo parameters $(Q, b_{\rm halo}, \rho_0, i)$ of our mass models by comparing with observational data, we employ a simple $\chi^2$ test,
\begin{eqnarray}
\chi^2 = \sum^{N_{\rm bins}}_{i} \frac{[\sigma^{\rm obs}_{i} - \sigma^{\rm model}_{i}(Q, b_{\rm halo}, \rho_0, i)]^2}{\epsilon^{2}_{i}},
\label{eq:eq18}
\end{eqnarray}
where $N_{\rm bins}$ is the number of bins, $\sigma^{\rm obs}$ is the measured velocity dispersion, $\sigma^{\rm model}$ is the predicted dispersion at the same distance, and $\epsilon$ is the uncertainly on $\sigma^{\rm obs}$. 
We here employ the reduced $\chi^2$ statistics which gives the value of $\chi^2/\nu$, where $\nu$ is the number of degree of freedom, usually given by $\nu = N_{\rm bins} - N_{\rm parameters}$, where $N_{\rm parameters}$ is the number of parameters. The results with reduced-$\chi^2$ values being near unity indicate acceptable fit, whereas these values being much lager than unity indicate that model fitting is not acceptable.

\section{RESULTS} 
\begin{deluxetable*}{cccccccc}
\label{tb:tab2}
\tablecolumns{8}
\tablewidth{6in}
\tablecaption{Results of $\chi^2$ fitting for six dSph galaxies and estimated mass within 300pc.}
\tablehead{
Galaxy & Halo Model & reduced-$\chi^2$ & $Q$ & $b_{\rm halo}$[pc] & $\rho_{0}[{\rm M}_{\odot}{\rm pc^{-3}}]$  & $i$ (inclination)[deg] & ${\rm M_{300}} {\rm [10^6M_{\odot}}]$ }
\startdata
Carina     &     NFW      &  1.90  &   $0.34\pm0.03$ & $1372\pm58$ & $0.0068\pm0.004$ & $71\pm3$ &  $1.75^{+0.36}_{-0.22}$ \\ 
               &     CORE    &  0.88  &   $0.39\pm0.03$   & $453\pm21$ & $0.065\pm0.003$ & $79\pm7$ & $2.05^{+0.23}_{-0.29}$ \\
Fornax    &     NFW      &  1.22  &   $0.42\pm0.03$ & $1823\pm51$ & $0.0072\pm0.0003$ & $74\pm3$ &  $3.07^{+0.41}_{-0.35}$ \\ 
               &     CORE    &  0.62  &   $0.37\pm0.02$   & $1323\pm53$ & $0.025\pm0.001$ &  $90_{-10}$ &  $1.03^{+0.10}_{-0.10}$\\
Sculptor  &     NFW      &  1.50  &   $0.68\pm0.04$ & $687^{+20}_{-17}$ & $0.031\pm0.001$ &  $90_{-10}$ &  $7.50^{+0.85}_{-0.87}$ \\ 
               &     CORE    &  1.08  &   $0.51\pm0.03$   & $405\pm16$ & $0.13\pm0.005$  & $84\pm6$ &  $5.01^{+0.53}_{-0.59}$ \\
Sextans  &     NFW      &  2.91  &   $0.41^{+0.06}_{-0.03}$ & $3510^{+207}_{-242}$ & $0.001\pm0.0001$  & $90_{-14}$ &  $0.81^{+0.27}_{-0.15}$ \\ 
               &     CORE    &  1.81  &   $0.31\pm0.02$   & $2880^{+580}_{-430}$ & $0.005\pm0.0004$  & $80^{+10}_{-4}$ &  $0.17^{+0.03}_{-0.02}$ \\
Draco     &     NFW      &  1.16  &   $0.39^{+0.08}_{-0.04}$ & $901^{+60}_{-70}$ & $0.029\pm0.003$  & $70^{+7}_{-5}$ &  $5.37^{+2.18}_{-1.36}$ \\ 
              &     CORE    &  1.14  &   $0.40\pm0.06$   & $359\pm30$ & $0.21\pm0.02$  & $80\pm10$ &  $5.81^{+1.98}_{-1.52}$ \\
LeoI       &      NFW      &  0.46  &   $0.90^{+0.14}_{-0.08}$ & $340\pm20$ & $0.11\pm0.01$ &  $80^{+10}_{-15}$ &  $14.2^{+5.10}_{-3.59}$ \\ 
              &      CORE    &  0.49  &   $0.41^{+0.05}_{-0.07}$   & $256^{+16}_{-14}$ & $0.35\pm0.04$ & $50\pm2$ &  $7.18^{+2.41}_{-1.93}$ 
\enddata
\end{deluxetable*}

\subsection{Best fit models of dark halos}
 \begin{figure}[t!]
\begin{center}
  \includegraphics[scale=0.7]{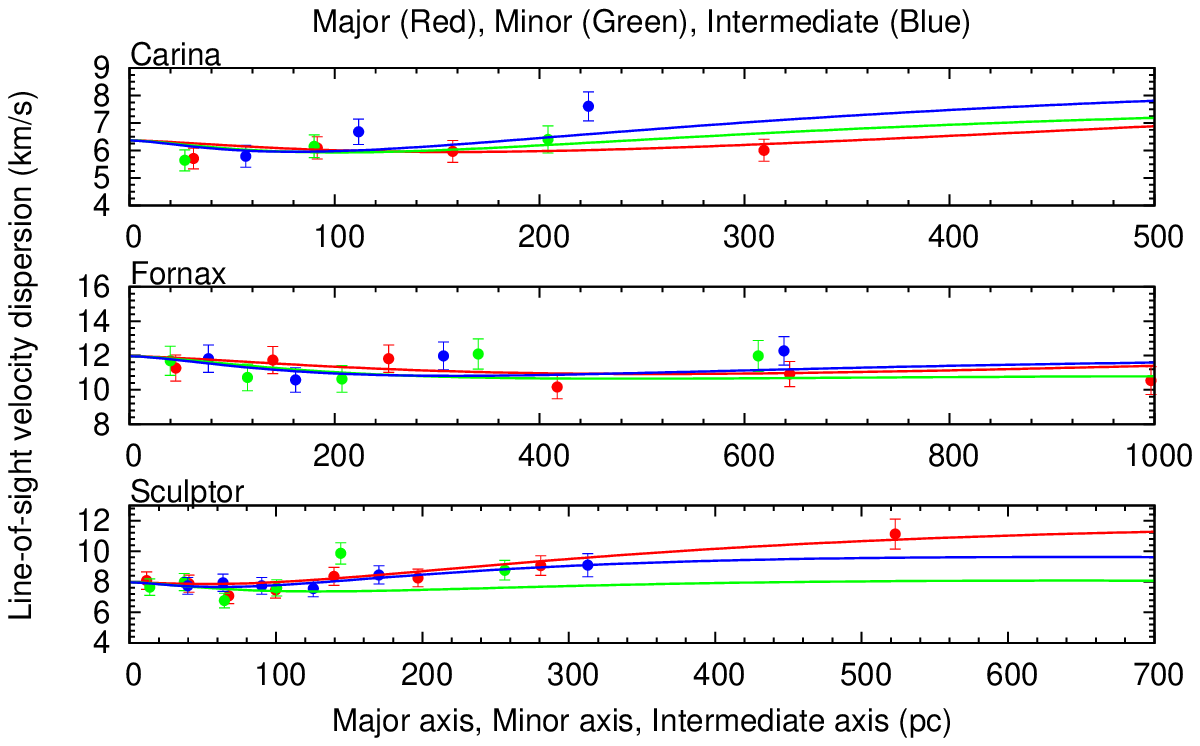}
 \end{center}
\end{figure}
 \begin{figure}[t!]
\begin{center}
  \includegraphics[scale=0.7]{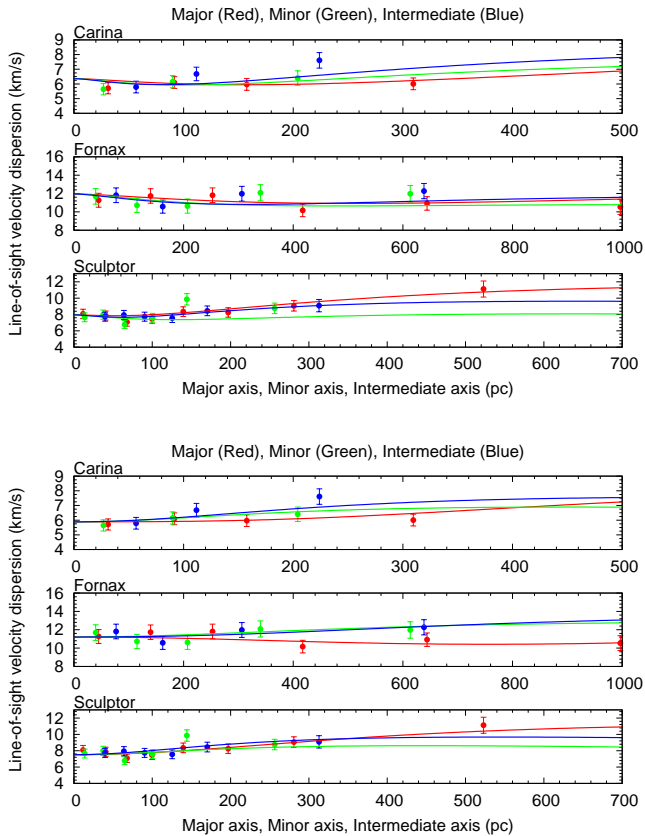}
 \end{center}
 \caption{The best-fit profiles of line-of-sight velocity dispersion along major, minor and intermediate axes for Carina, Fornax and Sculptor. The upper three panels are the case of NFW model, while the lower panels are the case of CORE model. Red, green and blue marks denote observed line-of-sight velocity dispersions along major, minor and intermediate axes, respectively. Red, green, blue lines are corresponding best-fit model results along the respective axes.}
 \label{fig:fig3}
\end{figure}

In this work, we focus on the shape of dark halos for two characteristic density profiles, centrally cusped and cored profiles. We confine ourselves to an NFW profile with $(\alpha, \delta) = (-1, -1)$ (referred to as NFW) and cored profile with $(\alpha, \delta) = (0, -1.5)$ (referred to as CORE), for both of which outer density profile is the same, $\rho(m) \propto m^{-3}$. 
Table 2 tabulates the best fit results for the halo parameters that we obtain from $\chi^2$ test for each of  six dSph satellites. Figures \ref{fig:fig3} and \ref{fig:fig4} show the best fit profiles of line-of-sight velocity dispersions along major, minor and intermediate axis for NFW (upper panel) and CORE (lower panel) models, respectively. 
It is clear from column 4 in Table 2 that the shapes of dark matter halos are generally not spherical, $Q \neq 1$, but oblate and flattened with $Q < 1$, both for NFW and CORE models. These results can be easily understood as follows. As described in the previous section, when a stellar system is flattened (as observed) for an assumed spherical halo, the shape of velocity dispersion profile is characterized by a wavy feature: trough and crest in the inner and outer parts, respectively.  However, velocity dispersion profiles obtained from the observational data appear to be almost flat for observed flattened stellar systems, thus it is obvious that dark halos are expected to be flattened. By contrast, as we will discuss later, Leo~I dSph appears to have a spherical dark halo. 

The relation between other parameters of dark halos is shown in Figure \ref{fig:fig5}. Figure \ref{fig:fig5}a shows the relation between scale lengths of dark halo and stellar component, $b_{\rm halo}$ and $b_{\ast}$. We find larger $b_{\ast}$ for larger $b_{\rm halo}$. Hence we estimate the ratio of two scale lengths, $b_{\rm halo}/b_{\ast}=2\sim3$, suggesting that a dark halo has a larger dimension than a stellar component and that the Galactic dwarf satellites are dominated by dark matter. On the other hand, we find that a scale density of dark halos, $\rho_0$, is smaller for a larger scale length, $b_{\rm halo}$, as shown in Figure \ref{fig:fig5}b. In particular, Sextans has a low density in both luminous and dark components. Therefore relatively compact dSphs have high density, while large dSphs are characterized by low density.
 \begin{figure}[t!]
\begin{center}
  \includegraphics[scale=0.7]{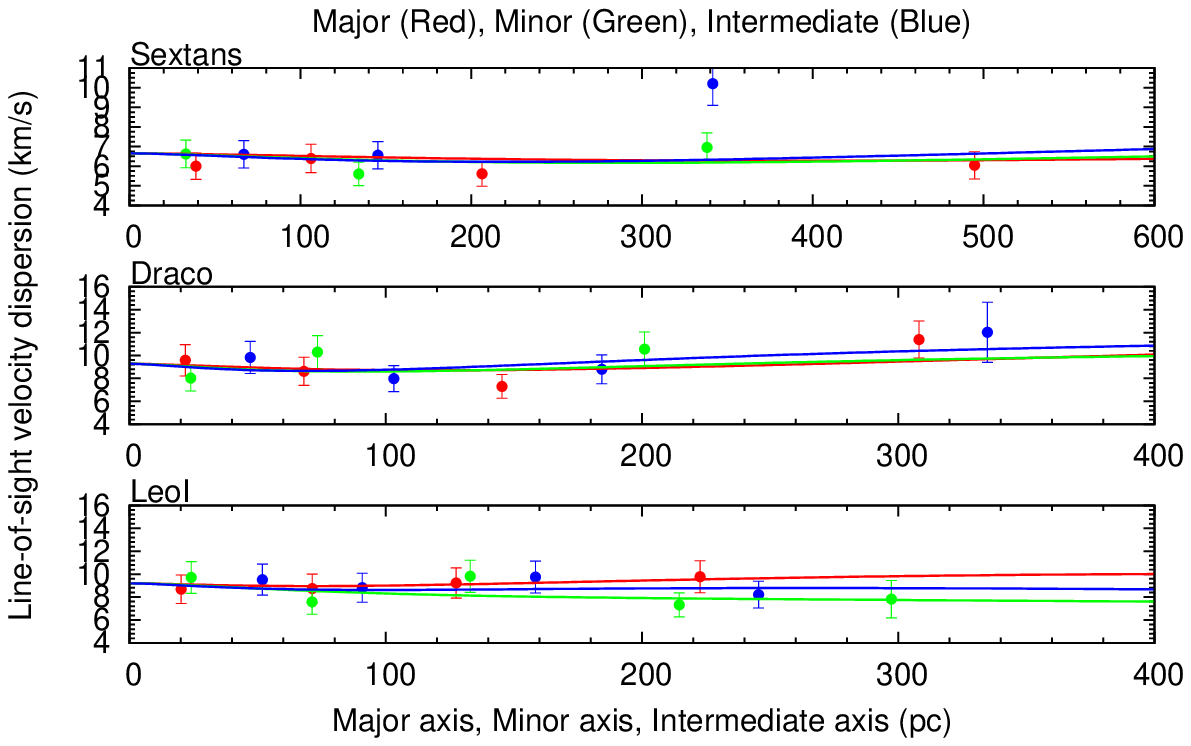}
 \end{center}
\end{figure}
 \begin{figure}[t!]
\begin{center}
  \includegraphics[scale=0.7]{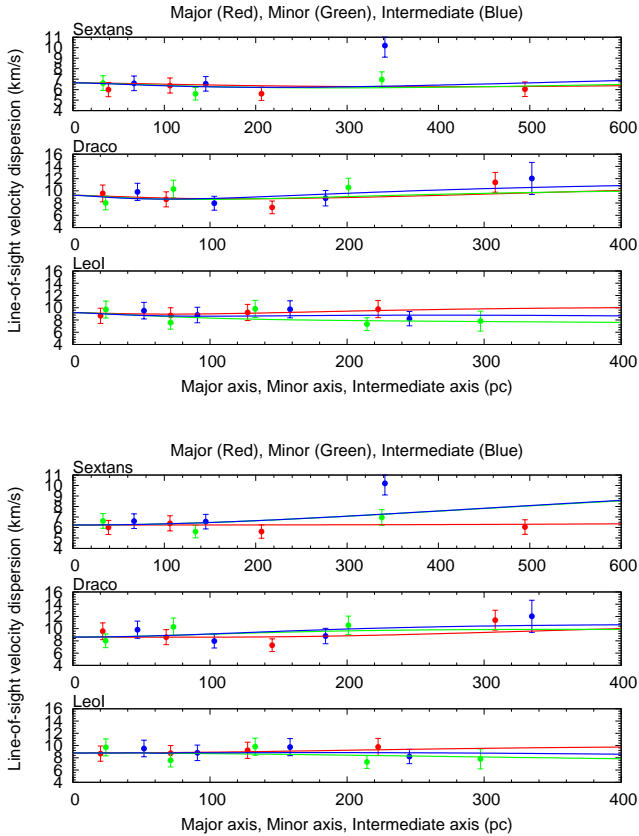}
 \end{center}
 \caption{The same as Figure \ref{fig:fig3} but for Sextans, Draco and Leo~I.}
 \label{fig:fig4}
\end{figure}

We stress here that our axisymmetric models are capable of taking into account the change of velocity anisotropy with the spatial coordinates. This is important because in order to understand the observed profile of line-of-sight velocity dispersion, $\sigma_{\rm los}$, it is needed to consider the effects of such a velocity anisotropy profile as well as those of flattened stellar and dark matter components, as we mentioned in \S 2.4. 
Figure \ref{fig:fig6} shows the ratio between $\sigma^2(= \overline{v^2_R}=\overline{v^2_z})$ and $\overline{v^2_{\phi}}$ for six dSphs along the major axis from the center to the farthermost data of each galaxy, as obtained from CORE model. For instance, for Carina (thick solid line) and Draco (thin dashed line), $\overline{v^2_{\phi}}$ is somewhat smaller than $\sigma^2$ in their inner parts. This velocity anisotropy can be understood as follows. As shown in Figure \ref{fig:fig3} and \ref{fig:fig4}, the line-of-sight velocity dispersion profiles of these galaxies are almost flat along the major axis, although their stellar systems are highly flattened, resulting in the non-flat profile of $\sigma_{\rm los}$ as shown in Figure \ref{fig:fig2}. Thus, the best-fit model is that the $\sigma_{\rm los}$ profiles at the inner parts be reproduced by radial anisotropy $\sigma^2 > \overline{v^2_{\phi}}$ and $Q<1$, while outer parts be tangentially anisotropic,  $\overline{v^2_{\phi}}>\sigma^2$, so that the resulting $\sigma_{\rm los}$ profiles are almost flat. For Fornax (thick dashed line) and Sextans (thin solid line), their $\sigma_{\rm los}$ profiles are decreasing with radii as shown in Figure \ref{fig:fig3} and \ref{fig:fig4}, respectively. Thus their velocity dispersion should be radially anisotropic, as shown in Figure \ref{fig:fig6}. On the other hand, velocity anisotropy in Sculptor (thick dotted-dashed line) is largely tangential because the $\sigma_{\rm los}$ profile is increasing with radius. By contrast, although the $\sigma_{\rm los}$ profile in Leo~I is nearly flat, velocity dispersion is tangentially anisotropic (thin dotted-dashed line). This arises from its rather low inclination angle ($i=50$ deg as shown in Table 2); low inclination suggests that intrinsic stellar axial ratio, $q$, is much small compared to projected one, $q^{\prime}$, and that the contribution of $\overline{v^2_{\phi}}$ to $\sigma_{\rm los }$ is less significant. Therefore, to reproduce the flat $\sigma_{\rm los}$ profile, $\overline{v^2_{\phi}}$ ought to be rather large in most parts. It follows from these studies that velocity anisotropy is not constant and universal in each dSph galaxy.

Finally, we investigate the degeneracy in model fitting for determining these four parameters $(Q,b_{\rm halo}, \rho_0,i)$. In Figure \ref{fig:fig7}, we present 68 \% (1ƒÐ), 95 \%, 99 \% confidence levels of contours in the two-dimensional plane of $Q-b_{\rm halo}$, $Q-b_{\rm halo}$, $Q-\rho_0$ and $b_{\rm halo}-\rho_0$ for CORE model of Fornax dSph. We confirm that we obtain the comparable results for NFW model and other dSphs as well (except for Sextans, as shown below). It follows that $Q-b_{\rm halo}$, $Q-\rho_0$ and $Q-i$ contour maps show little degeneracy with respect to $Q$, therefore we can determine the shapes of dark halos without strong degeneracies within the framework of axisymmetric mass models. Moreover, we can determine $b_{\rm halo}$ and $\rho_0$ independent of inclination, so an inclination angle, $i$, has little influence on the determination of other halo parameters. Contrary to this, as seen from $b_{\rm halo}$-$\rho_0$ map, there exists an obvious degeneracy between these two parameters; both of these parameters affect the total amplitude of velocity dispersions. Thus, with available data alone it is difficult to break this $b_{\rm halo}$-$\rho_0$ degeneracy.

In contrast to Fornax and other dSphs, Sextans yields rather high degeneracies in $Q-b_{\rm halo}$ and also $Q-\rho_0$ as shown in Figure \ref{fig:fig8}. Thus, although a nominal $\chi^2$ fitting provides the smallest $Q$~$(=0.31)$ for one model (CORE) of this galaxy, a yet higher $Q$ of 0.4 to 0.5 would be allowed depending on model specifications. 

\begin{deluxetable*}{cccccc}[hb!!!]
\label{tb:tab3}
\tablecolumns{6}
\tablewidth{5in}
\tablecaption{Comparison of $\chi^2$ fitting for six dSph galaxies when $Q$ is fixed.}
\tablehead{
Galaxy & Halo Model & \multicolumn{4}{c}{reduced-$\chi^2$}\\
            &                    & $Q=1$ & $Q=0.9$  & $Q=0.7$  & Best fit case
            }
\startdata
Carina     &     NFW      &  5.81  & 5.22   & 3.74  & 1.90 ($Q=0.34$)\\ 
               &     CORE    &  9.32\tablenotemark{a}  & 7.69   & 4.11  & 0.88 ($Q=0.39$)\\
Fornax    &     NFW      &  4.12\tablenotemark{a}  & 3.61   & 2.73  & 1.22 ($Q=0.42$)\\ 
               &     CORE    &  5.21\tablenotemark{a}  & 3.69\tablenotemark{a}   & 3.06\tablenotemark{a}  & 0.62 ($Q=0.37$)\\
Sculptor  &     NFW      &  2.37  & 2.03   & 1.41  & 1.50 ($Q=0.68$)\\
               &     CORE    &  3.84\tablenotemark{a}  & 2.39\tablenotemark{a}   & 1.57  & 1.08 ($Q=0.51$)\\
Sextans  &     NFW      &  3.27  & 3.06   & 2.66  & 2.91 ($Q=0.41$)\\
               &     CORE    &  3.69\tablenotemark{a}  & 3.25\tablenotemark{a}   & 2.61  & 1.81 ($Q=0.31$)\\
Draco     &     NFW      &  2.22  & 1.99   & 1.45  & 1.16 ($Q=0.39$)\\
              &     CORE    & 3.16   & 2.63   & 1.75  & 1.14 ($Q=0.40$)\\
LeoI       &      NFW      & 0.40  & 0.41   & 0.42  & 0.46 ($Q=0.90$)\\ 
              &      CORE    & 0.51\tablenotemark{a}  & 0.37\tablenotemark{a}   & 0.40\tablenotemark{a}  & 0.49 ($Q=0.41$)
\enddata
\tablenotetext{a}{These cases yield scale lengths of dark halos, $b_{\rm halo}$, shorter than those of stellar components, $b_{\ast}$, so may be unrealistic.}
\end{deluxetable*}

\subsection{Cusped or cored dark halo}
We now address the question which central density profile, cusped or cored, is acceptable in light of observed velocity data. From our fitting results we find that a cored profile of dark halos (CORE) fits somewhat better than cusped one (NFW) (see column 3 in Table 2). In particular, for Carina, Fornax, Sculptor and Sextans dSphs, we find that CORE model accords rather well with data based on $\chi^2$ fitting. Recent studies on this issue have suggested the following two different results. Walker et al. (2009c) used kinematic data obtained with the Michigan/MIKE Fiber System at the Magellan/Clay 6.5 m telescope (see Walker et al. 2007), and they applied the spherical Jeans equation to estimate masses for eight of the brightest dSphs. They found that kinematic data match both NFW and CORE, thus cusp-core problem has remained an issue. In contrast, Strigari et al. (2010) assumed that the surface brightness of dSphs has shelving cusps and that instead of assuming NFW, they made use of the density profiles of the subhalos which are directly obtained from Aquarius simulation in the framework of $\Lambda$CDM theory. Their model fitting based on spherical Jeans equation reproduced observational data, so they concluded that current data on the Galactic satellites are consistent with the hypothesis that these galaxies live in $\Lambda$CDM halos. Although the clear solution to the cusp-core problem is yet unavailable, our work based on more realistic, non-spherical mass models appears to support a cored central density in these dSphs. 
 \begin{figure}[t!]
\begin{center}
  \includegraphics[scale=0.7]{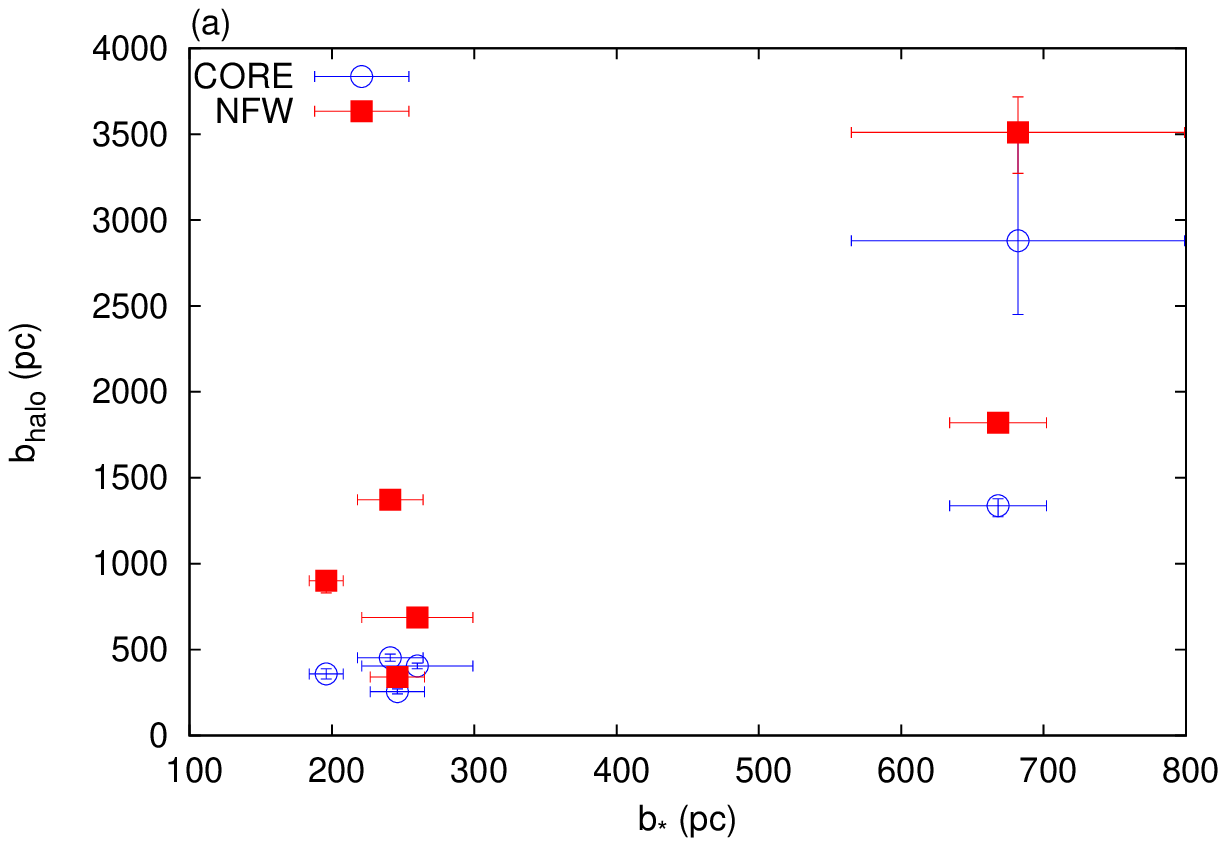}
  \includegraphics[scale=0.7]{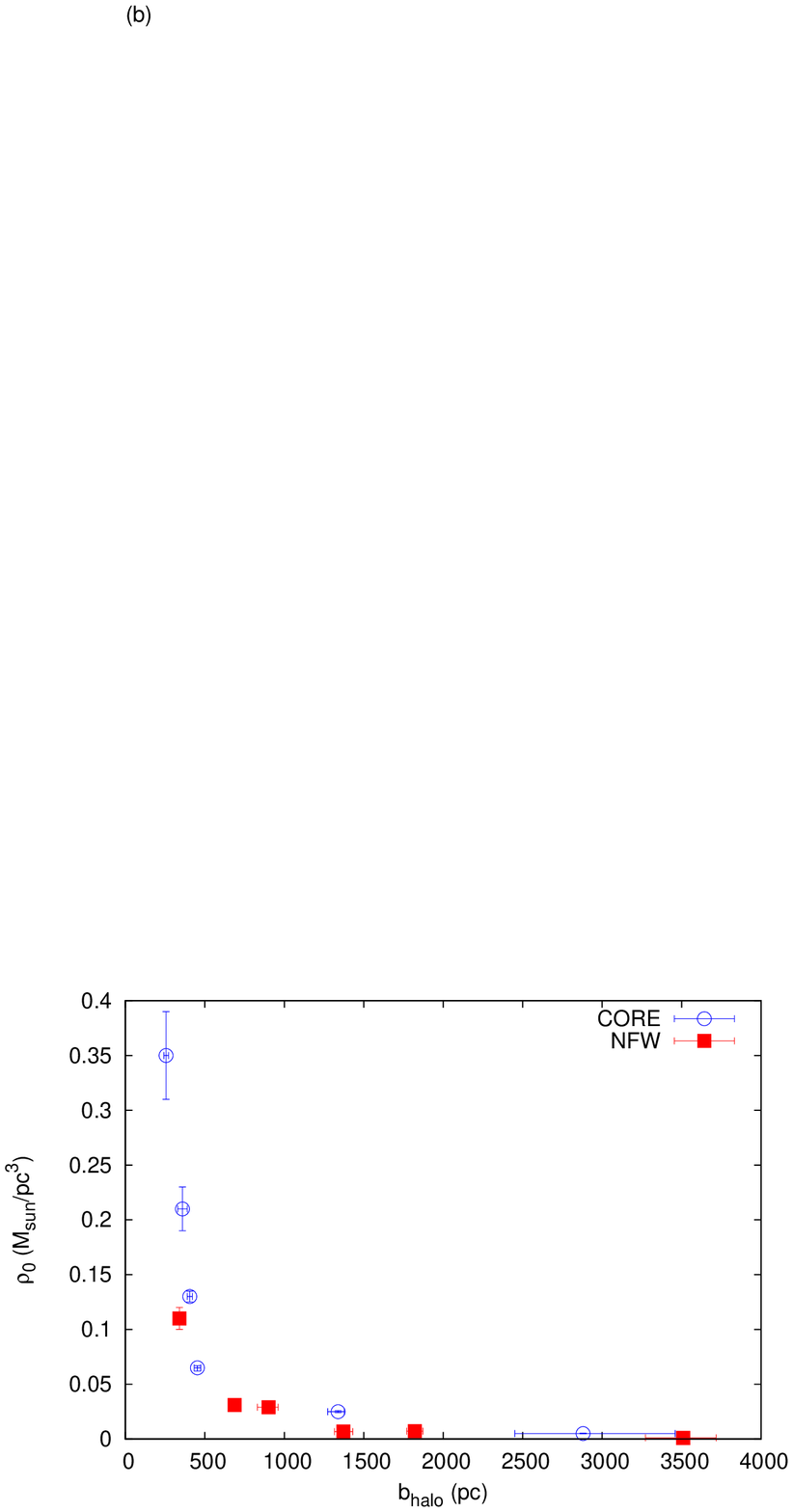}
 \end{center}
 \caption{The relation between $Q$ and $b_{\rm halo}$ (upper panel) and between $\rho_0$ and $b_{\rm halo}$ (lower panel) for six dSphs obtained from CORE and NFW model.}
 \label{fig:fig5}
\end{figure}

However, it is worth noting that even if the dark matter is largely dominated in a gravitational potential, the specific assumption of stellar density profiles can affect the inferred slope of total density profiles at inner parts of a dSph (Evans, An \& Walker 2009). To investigate this issue, we consider another functional form for the stellar component, namely an exponential density profile, $\nu(R,z) = \nu_0\exp(-m_{\ast}/b_{\ast})$ where $m^2_{\ast} = R^2 + z^2/q^2$, and $\nu_0$ and $b_{\ast}$ are scale density and length, respectively. We adopt the half light radius for $b_{\ast}$ (Irwin \& Hatzidimitriou 1995) and perform the same velocity analysis as for a Plummer model. 
Taking the case of Fornax dSph as an example, we obtain $\chi^2_{\nu}=1.59$ $(0.97)$ for NFW (CORE) model, while a Plummer stellar density yields $\chi^2_{\nu}=1.22$ $(0.62)$ for NFW (CORE). Thus there is little significant difference in $\chi^2$ values between Plummer and exponential models for the stellar component and also between NFW and CORE models, so it remains unclear which central density profile of dark halos is preferred from observations. 
We note that the best-fit shapes of dark halos using an exponential stellar profile are slightly rounder than those using a Plummer one: the former yields $Q=0.48$ $(0.38)$ for NFW (CORE) compared to $Q=0.42$ $(0.37)$ for the latter (Table 2). We obtain the similar results for other dSphs as well. This is because exponential three-dimensional profiles are somewhat steeper in the central parts than Plummer ones, thereby increasing the inner part of the $\sigma_{\rm los}$ profile, as is the case for the difference between NFW and CORE on $\sigma_{\rm los}$. Thus to reproduce a rather flat feature of the observed $\sigma_{\rm los}$ profile, the best fit $Q$ be slightly larger for an exponential stellar profile. We thus suggest that systematic uncertainties for $Q$ values by adopting different mass or stellar density profiles are limited only to about $0.1$.
\begin{figure}
\figurenum{6}
\begin{center}
  \includegraphics[scale=0.7]{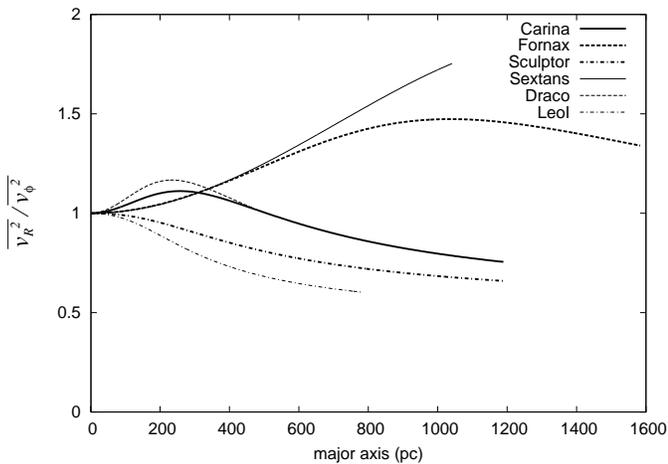}
\end{center}
\caption{The ratio $\overline{v^2_R}/\overline{v^2_{\phi}}$ for best-fit CORE models of six dSphs along the major axis (up to the far end of the observational data). Thick solid, dashed and dotted-dashed lines are for Carina, Fornax and Sculptor, while thin three lines are for Sextans, Draco, and Leo~I, respectively. The label in the vertical axis denote $\overline{v^2_R}/\overline{v^2_{\phi}}$ with the assumption $\overline{v^2_z}=\overline{v^2_R}$.}
\label{fig:fig6} 
\end{figure}

\subsection{Flattened or spherical dark halo}
\begin{figure*}
\figurenum{7}
\plotone{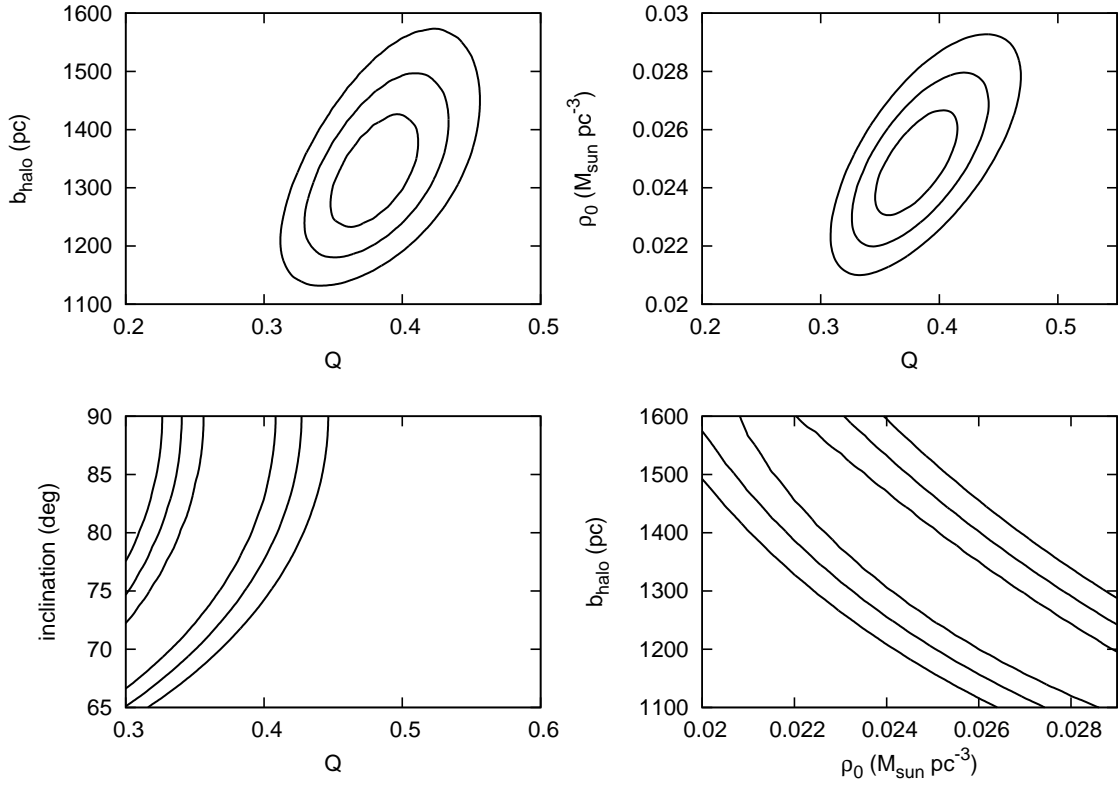}
\caption{Likelihood contours for each dark halo parameter for Fornax dSph based on our CORE models. Contours show 68\% ($1\sigma$), 95\%, and 99\% confidence levels. Clockwise from top left, $Q-b_{\rm halo}$, $Q-\rho_0$, $\rho_0-b_{\rm halo}$, $Q-i$ contours.}
\label{fig:fig7} 
\end{figure*}

\begin{figure*}
\figurenum{8}
\plotone{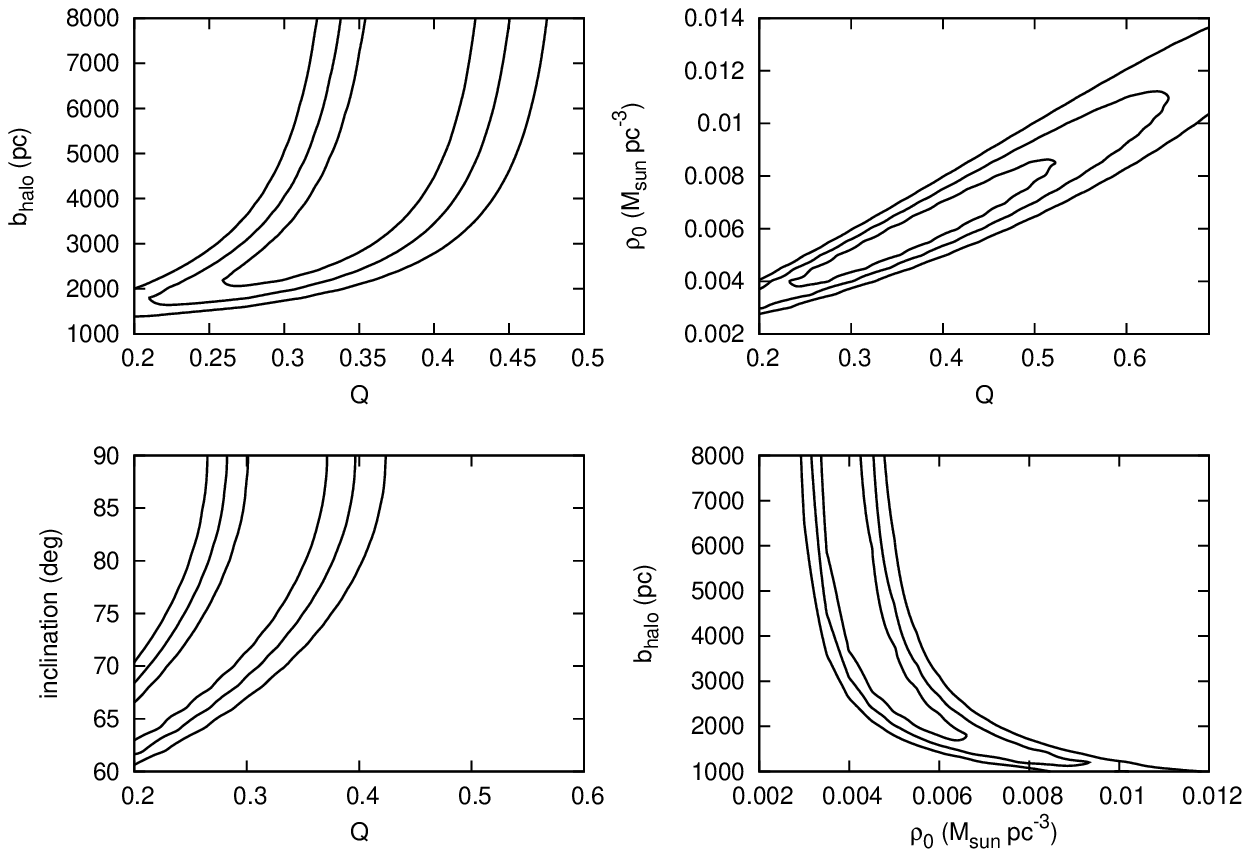}
\caption{The same as Figure \ref{fig:fig7} but for the case of Sextans.}
\label{fig:fig8} 
\end{figure*}

We set further limits on the flattened shape of dark halos in comparison with spherical shape. In particular, visible parts of dSphs are actually non-spherical, that is, stellar distributions are flattened with $q<1$, thus we compare fitting results (with this setting of $q<1$) between $Q=1$ and $Q<1$ cases.
Table 3 shows the results of $\chi^2$ fitting when we fix $Q=1$. Reduced-$\chi^2$ values of most galaxies are much lager than those for $Q<1$ shown in Table 2. We note that the best fit model with $Q=1$ shows $b_{\rm halo}<b_{\ast}$ in Fornax, Sextans for CORE and other dSphs, which is unrealistic. Thus spherical halo models are inconsistent with most of observed dSphs.

Table 3 also shows the cases of $Q=0.9$ and $0.7$, which approximate $\Lambda$CDM prediction for likely intermediate-to-major and minor-to-major axial ratios for subhalos, respectively, as described in more detail in \S5.2. It is found that the reduced-$\chi^2$ values for the sample dSphs (except for Leo~I) are smaller than those with $Q=1$, but they are still much larger than unity so the case of $Q=0.9\sim0.7$ does not adequately reproduce the data of these dSphs.

It is worth noting that, for Leo~I, spherical NFW models can be fitted to observational data reasonably well. This may be explained as follows.
Since Leo~I is the most distant from the center of the Galaxy among six dSphs, the effect of tidal force on Leo~I may be the smallest. Additionally, according to N-body simulations, less massive dark matter halos tend to be more relaxed than massive halos and thus are more spherical (Wang et al. 2011; Ragone-Figueroa et al. 2010; Schneider et al. 2011). 
In order to confirm these suggestions, we require reliable proper motion data of Leo~I and obtain its accurate orbital motion to examine the role of Galactic tides in the shape of its dark halo.

\section{DISCUSSION}
Based on our new mass limits on the Galactic dSphs, we discuss the detailed properties of dark halo structures obtained here in comparison with previous mass limits and $\Lambda$CDM predictions.

\subsection{Mass within 300 parsecs}
Recently, Strigari et al. (2008, hereafter S08) claimed that all the dSph satellites with luminosity over almost four orders of magnitude have a common mass of $\sim10^7M_{\odot}$ within a radius of 300 pc ($M_{300}$), where dark matter is dominated in its mass. This result suggests particular restriction on the properties of dark matter and the formation process of dwarf galaxies; several attempts have been made to reproduce a common mass scale of around $10^7M_{\odot}$. For example, Macci\`o et al. (2009) have performed numerical simulations coupled with a semi-analytical model for galaxy formation, to explain common mass scale within the context of $\Lambda$CDM scenario. 
They have suggested that the narrow range of $M_{300}$ originates from the narrow distribution of circular velocities ($V_{\rm circ}=20\sim40$ km~s$^{-1}$) of the progenitor subhalos at the time of their accretion on to a host halo, for which baryonic matter is able to cool rapidly and form stars.
In this model the wide range of satellite luminosities is a combination of mass at time of accretion and the broad distribution of accretion redshifts for a given mass. 

However, S08 have assumed spherical symmetry for both stellar and dark halo density profiles and solve spherically symmetric Jeans equation for simplicity. We thus compare the results of these spherical mass models with those of the current work. 
We use equation (\ref{eq:eq10}) to estimate $M_{300}$, which, in this work, is defined as mass enclosed within a spheroid with major-axis length of 300~pc and axial ratio $Q$.
Figure \ref{fig:fig8} illustrates the estimated total mass within 300 pc as a function of their total luminosity. The filled symbols denote the spherical mass model, while open symbols are based on axisymmetric models for the cases of NFW (triangles) and CORE (circles). It is clear that our axisymmetric mass models provide a different picture on this issue, namely the mass constancy within inner 300 pc as argued by spherical models is not necessarily the case. Therefore, we find that this mass estimate is rather sensitively dependent on the assumed mass profiles and shapes of dark matter halos in dSphs.
Whether or not this is also the case for less luminous dSphs is yet unclear because of a small number of available sample stars.

\subsection{Comparison with the $\Lambda$CDM models}
As mentioned in \S 1, $\Lambda$CDM theory has yet several discrepancies with existing observations on the spatial scales smaller than $\sim1$~Mpc, i.e., galactic and subgalactic scales. Recently a new issue has been raised that the masses of most massive subhalos in a galaxy sized halo in $\Lambda$CDM are systematically heavier than those of the Galactic satellites (Boylan-Kolchin et al. 2012, hereafter BK12; Ferrero et al. 2011). We assess this issue from our current mass models, following the procedure of BK12. They compared maximum $V_{\rm circ}$, $V_{\rm max}$, of most massive ten subhalos in galaxy-sized halos in $\Lambda$CDM with those of brightest dSphs. While all dSphs have $12 \lesssim V_{\rm max} \lesssim 25$ km~s$^{-1}$, ten subhalos predicted by $\Lambda$CDM simulation have $V_{\rm max}>25$ km~s$^{-1}$ (See Figures 2 and 6 of BK12). Using axisymmetric models, we also calculate $V_{\rm max}$ of the six dSphs which we have employed here. 
The circular velocity can be calculated
\begin{equation}
V^2_{\rm circ}(R) = R|-\nabla\Phi|,
\label{eq:eq19}
\end{equation}
where $\nabla\Phi$ can be easily derived from equation (\ref{eq:eq5}). We estimate $V_{\rm max}$ along the major axis, $R$, using
\begin{equation}
g_{R}(R,z) = -\frac{\partial \Phi}{\partial R} = -2\pi GQa_0^3R\int^{\infty}_{0}d\tau\frac{\rho(R,z)}{(\tau+a_0^2)^2\sqrt{\tau+Q^2a_0^2}}.
\label{eq:eq20}
\end{equation}
We find that our estimated $V_{\rm max}$ from axisymmetric models is somewhat smaller than spherical models: we obtain $V_{\rm max} = 10.4\pm1.1$ km~s$^{-1}$ for Carina, $18.8\pm1.6$ km~s$^{-1}$ for Fornax, $14.5\pm1.1$ km~s$^{-1}$ for Sculptor, $16.6\pm4.7$ km~s$^{-1}$ for Sextans, $14.9\pm2.9$ km~s$^{-1}$ for Draco and $13.8\pm2.4$ km~s$^{-1}$ for Leo~I. We confirm the claim by BK12 that observed $V_{\rm max}$ values of the Galactic dSphs are systematically smaller than those of $\Lambda$CDM subhalos having $V_{\rm max} > 25$ km~s$^{-1}$.

We further test the predictions of $\Lambda$CDM theory, using derived shape distributions of dSphs' dark halos. In Schneider et al. (2011), the distribution of axial ratios of triaxial CDM halos in mass scales ($10^{9.8}h^{-1}M_{\odot}\leq M \leq 10^{14.3}h^{-1}M_{\odot}$, where $h = H_0/100$ km~s$^{-1}$~Mpc$^{-1}$) were derived from the Millennium and Millennium-2 dark matter N-body simulations (Springel et al. 2005; Boylan-Kolchin et al. 2009). It is found that less massive halos are more spherical. A likely explanation for this is that because of more survived substructures in more massive halos, such halos tend to be less relaxed and are therefore less spherical than less-massive halos (See Figure 3 of Schneider et al. 2011). We now focus on less-massive halos ($10^{9.8}h^{-1}M_{\odot}\leq M \leq 10^{10.3}h^{-1}M_{\odot}$) in the simulations, since  this mass range covers the mass of typical subhalos employed here. The distribution of axial ratios of these low mass halos shows a peak at about 0.9 for intermediate-to-major axial ratio and 0.7 for minor-to-major axial ratio, and the probability is very small at axial ratios less than 0.5. On the other hand, the axial ratio of dark halos in the Galactic satellites obtained by our axisymmetric mass model is in the range of $0.3 \leq Q \leq 0.5$ for CORE models, which are smaller than theoretically predicted ones, even taking into account possible systematic uncertainties of about $\Delta Q=0.1$ by using different stellar density profiles. 
It is also noted that as shown in \S4.2, reduced-$\chi^2$ values in the fitting with fixed $Q$ of 0.9 and 0.7 are much larger than unity and thus unacceptable.
We further compare with results of Via Lactea-I simulation (Kuhlen et al. 2007), in which radially-dependent axial ratios for subhalos are presented (See their Figure 3). We estimate the mass weighted average of axial ratios given in their Figure 3 to compare with $Q$ in our mass models. Specifically, provided that subhalos obtained from N-body simulation have NFW density profile, we calculate mass weighted average of major-to-intermediate axial ratios (supposing $a\ge b\ge c$) 
\begin{equation}
\Bigl<\frac{b}{a}\Bigr> = \frac{\int^{R_{t}}_{0} \Bigl(\frac{b}{a}\Bigr)_{r}\rho dV}{\int^{R_{t}}_{0}\rho dV} \hspace{5mm} (dV=4\pi r^2dr),
\label{eq:eq21}
\end{equation}
where $R_{t}$ is the radius of subhalos and $\rho$ is an assumed NFW density profile. $<c/a>$ is calculated in the same manner. As a result, we obtain $<b/a>=0.8\pm0.1$ and $<c/a> = 0.64\pm0.1$, thus these axial ratios are much larger than our results. 
Therefore, the shapes of subhalos predicted from $\Lambda$CDM-based N-body simulations are inconsistent with observations of the Galactic dwarf satellites.

It is interesting to remark that our $Q$ values for the Galactic dSphs are generally in agreement with the axial ratios measured for mass distributions of elliptical galaxies using their X-ray halos (e.g., Buote \& Canizares 1994) and/or strong gravitational lensing (e.g., Keeton, Kochanek \& Seljak 1997).

\subsection{Implications for dynamical evolution of subhalos}
The shapes of subhalos are generally subject to strength and frequency of tidal effects from the host halo (Kuhlen et al. 2007), thus if we find some relationship between the shapes of dark halos and the orbital parameters of dSphs in the Galaxy, we would be able to set useful limits on the dynamical evolution of subhalos. Assuming some specific model for the potential of the Galactic halo, we are able to estimate a pericentric distance and an orbital eccentricity from the position and proper motion of dSphs. Using these results, we investigate the possible relations between the axial ratio of dark halos, $Q$, and the orbital parameters of dSphs, but find no remarkable relations within uncertainties of these parameters. In this respect, more precise information of proper motions will be useful to obtain more precise orbital motions of dSphs and to set more useful constraints on the dynamical evolution of subhalos.

This work also suggests that the dark-matter structures of bright Milky Way satellites are not consistent with those of CDM subhalos in galaxy-sized host halos predicted by high-resolution N-body simulations: observed velocity distributions of stars in dSphs indicate a more flattened and less dense dark halo than predicted. There are several possible mechanisms to solve these discrepancies. For instance, baryonic feedback in dwarf satellites induced by supernovae explosion, namely rapid removal of substantial amounts of gas may cause a dynamical re-arrangement of mass distribution, thereby reducing the central density of dark matter. However, to what extent this effect actually works is yet unclear. Alternatively, CDM may differ its standard representations on small scales, e.g., replaced by Warm Dark Matter theory (Colin et al. 2000; Bode et al. 2001; Macci\`o et al. 2010; Lovell et al. 2012) or Self-Interacting Dark Matter theory (Moore et al. 2000; Spergel \& Steinhardt 2000; Yoshida et al. 2000), see the article by Ostriker \& Steinhardt (2003).
Even in the framework of CDM models, we point out a possible mechanism to preclude the above discrepancy, which has not been fully investigated in N-body simulations. In contrast to collisionless representation of a galaxy-sized host halo in many of N-body simulations, such a large halo actually contains a large amount of baryons, especially at its center so that the total matter distribution is approximated by an isothermal sphere rather than an NFW profile with shallower density slope. In other words, a host halo would have deeper and steeper density profile. Then subhalos passing through this potential may undergo more powerful tidal force than the case of a dark matter alone represented by NFW. Consequently, such tidally affected subhalos would be more diffuse by virtue of heavy mass loss, that is, $V_{\rm max}$ of subhalos can be reduced to observed values. Furthermore, the shapes of subhalos also may be more flattened by stronger tidal interaction with a host galaxy system. Although it is still speculative, more simulation work would be worth exploring based on this hypothesis.  
\begin{figure}
\figurenum{9}
\begin{center}
  \includegraphics[scale=0.7]{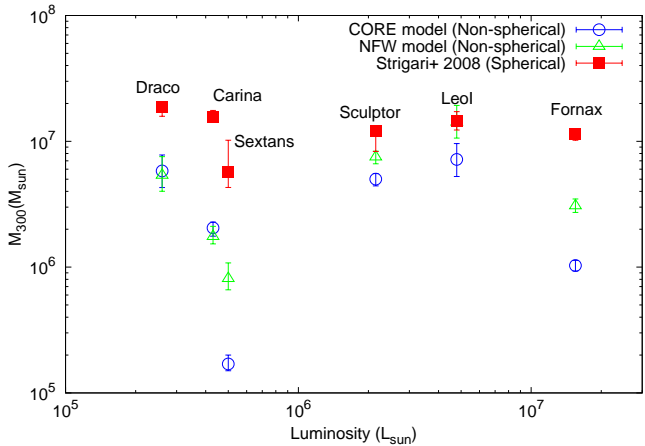}
\end{center}
\caption{The estimated total mass of six dSphs within their inner 0.3 kpc ($M_{300}$) as a function of their total luminosity, in units of solar luminosities. Filled squares denote the results of Strigari et al. (2008), while open circles (CORE) and triangles (NFW) are based on this work.}
\label{fig:fig9} 
\end{figure}

\section{Conclusion}
In this paper, we have constructed axisymmetric mass models for the dSphs in the Milky Way to obtain plausible limits on density profiles and shapes of their dark halos. This is motivated by the fact that most of mass models for the dSphs have assumed spherical symmetry for simplicity, despite the facts that the luminous parts of the dSphs are actually non-spherical and CDM models predict non-spherical virialized dark halos.
Our models consider velocity anisotropy of stars $\overline{v^2_R} / \overline{v^2_{\phi}}$, which can vary with the adopted cylindrical coordinates under the assumption $\overline{v^2_z}=\overline{v^2_R}$ for simplicity, and also include an inclination of the system as a fitting parameter to explain the observed line-of-sight velocity dispersion profile.
Based on the application of our models to six dSphs in the Galaxy, we have found that the best-fitting cases for most of the dSphs yield oblate and flattened dark halos, being independent of an assumed central density profile. We have calculated the total mass of the dSphs enclosed within a spheroid with major-axis length of 300~pc and found that the picture of the mass constancy within 300~pc as argued by spherical models is not necessarily supported. Recently, using axisymmetric Schwarzschild method, Jardel and Gebhardt (2012) applied their axisymmetric model to Fornax dwarf galaxy. They estimated the mass within inner 300~pc of $ M_{300} = 3.5^{+0.8}_{-0.1}\times10^6M_{\odot} $, which is in good agreement with our result of $M_{300} = 3.07^{+0.41}_{-0.35}\times 10^6M_{\odot}$ $(1.03^{+0.10}_{-0.10}\times 10^6M_{\odot})$ for NFW (CORE), but in contrast to $M_{300}= 1.14^{+0.09}_{-0.12}\times 10^7M_{\odot}$ obtained from a spherical model (Strigari et al. 2008). 
It is also found that the observed mass density of the dSphs, expressed in terms of $V_{\rm max}$, is smaller than that predicted by $\Lambda$CDM-based N-body simulations, as claimed by recent work. Furthermore, the axial ratios of the dSphs obtained here are systematically smaller than theoretical predictions. Further studies are needed to solve these issues.

\acknowledgments
The authors thanks the referee for her/his constructive comments that have helped us to improve our paper.
This work has been supported in part by a Grant-in-Aid for Scientific Research (20340039, 18072001) of the Ministry
of Education, Culture, Sports, Science and Technology in Japan and by JSPS Core-to-Core Program ``International Research Network for Dark Energy''.



\end{document}